\documentclass[natbib]{svjour3}                     % onecolumn (standard format)
\smartqed  % flush right qed marks, e.g. at end of proof
\usepackage{graphicx}
\usepackage{aps-bibstyle}  % use this style if you don't use BibTeX.
\newcommand{\aap}{{Astron. Astrophys.}}
\newcommand{\mnras}{{Monthy Notices R. Astron. Soc.}}
\newcommand{\araa}{{Ann. Rev. Astron. Astrophys.}}
\newcommand{\apj}{{Astrophys. J.}}
\newcommand{\apjl}{{Astrophys. J. Lett.}}
\newcommand{\prd}{{Phys. Rev. D}}
\newcommand{\pra}{{Phys. Rev. A}}
\newcommand{\prl}{{Phys. Rev. Lett.}}

\newcommand{\sm}[1]{\mbox{{\scriptsize #1}}}
\newcommand{\simle} {\,{}^<_{\sim}\,}

\newcommand{\bef}{\begin{figure}}
\newcommand{\eef}{\end{figure}}

\def\eps@scaling{0.95}

\def\showone#1{
  \centering
  \leavevmode
  \epsfxsize=\eps@scaling\linewidth
  \epsfbox{#1.eps}
}

\def\showtwover#1#2{
  \centering
  \leavevmode
  \epsfxsize=\eps@scaling\linewidth
  \epsfbox{#1.eps} \hfil
  \epsfxsize=\eps@scaling\linewidth
  \epsfbox{#2.eps}
}

\def\showthreeover#1#2#3{
  \centering
  \leavevmode
  \epsfxsize=\eps@scaling\linewidth
  \epsfbox{#1.eps} \hfil
  \epsfxsize=\eps@scaling\linewidth
  \epsfbox{#2.eps} \hfil
  \epsfxsize=\eps@scaling\linewidth
  \epsfbox{#3.eps}
}

\def\showfourover#1#2#3#4{
  \centering
  \leavevmode
  \epsfxsize=\eps@scaling\linewidth
  \epsfbox{#1.eps} \hfil
  \epsfxsize=\eps@scaling\linewidth
  \epsfbox{#2.eps} \hfil
  \epsfxsize=\eps@scaling\linewidth
  \epsfbox{#3.eps} \hfil
  \epsfxsize=\eps@scaling\linewidth
  \epsfbox{#4.eps}
}

\def\epstwo@scaling{0.46}

\def\showtwo#1#2{
  \centering
  \leavevmode
  \epsfxsize=\epstwo@scaling\linewidth
  \epsfbox{#1.eps}
  \epsfxsize=\epstwo@scaling\linewidth
  \epsfbox{#2.eps}
}

\def\epsthree@scaling{0.28}
\def\showthree#1#2#3{
  \centering
  \leavevmode
  \epsfysize=\epsthree@scaling\textwidth %% \linewidth
  \epsfbox{#1.eps}
  \epsfysize=\epsthree@scaling\textwidth %% \linewidth
  \epsfbox{#2.eps}
  \epsfysize=\epsthree@scaling\textwidth %% \linewidth
  \epsfbox{#3.eps}
}

\def\epstwo@scaling{0.44}

\def\showfour#1#2#3#4{
  \centering
  \leavevmode
  \epsfxsize=\epstwo@scaling\linewidth
  \epsfbox{#1.eps} \hfil
  \epsfxsize=\epstwo@scaling\linewidth
  \epsfbox{#2.eps} \hfil
  \epsfxsize=\epstwo@scaling\linewidth
  \epsfbox{#3.eps} \hfil
  \epsfxsize=\epstwo@scaling\linewidth
  \epsfbox{#4.eps}
}

\def\showsix#1#2#3#4#5#6{
  \centering
  \leavevmode
  \epsfxsize=\epstwo@scaling\linewidth
  \epsfbox{#1.eps} \hfil
  \epsfxsize=\epstwo@scaling\linewidth
  \epsfbox{#2.eps} \hfil
  \epsfxsize=\epstwo@scaling\linewidth
  \epsfbox{#3.eps} \hfil
  \epsfxsize=\epstwo@scaling\linewidth
  \epsfbox{#4.eps} \hfil
  \epsfxsize=\epstwo@scaling\linewidth
  \epsfbox{#5.eps} \hfil
  \epsfxsize=\epstwo@scaling\linewidth
  \epsfbox{#6.eps}
}

\newcommand{\befone}{
  \begin{figure*}
  \centering
  \begin{minipage}{\textwidth}
  }
\newcommand{\eefone}{\end{minipage}\end{figure*}}

\newcommand{\HI}{$\mathrm{H}$\ }

\newcommand{\HzI}{$\mathrm{H}_2$\ }

\newcommand{\HM}{$\mathrm{H}^-$\ }

\newcommand{\HeI}{$\mathrm{He}$\ }

\newcommand{\HDI}{$\mathrm{HD}$\ }

\newcommand{\HIId}{$\mathrm{H}^+$}
\newcommand{\HzIId}{$\mathrm{H}_2^+$}
\newcommand{\HzId}{$\mathrm{H}_2$}

\newcommand{\HMd}{$\mathrm{H}^-$}

\newcommand{\HeHIId}{$\mathrm{HeH}^+$}

\newcommand{\DId}{$\mathrm{D}$}
\newcommand{\DIId}{$\mathrm{D}^+$}
\newcommand{\HDIId}{$\mathrm{HD}^+$}

\newcommand{\DMd}{$\mathrm{D}^-$}

\newcommand{\fHeI}{\mathrm{He}}

\newcommand{\fe}{\mathrm{e}}
\newcommand{\fpp}{\mathrm{p}}

\begin{document}

\title{The First Magnetic Fields%\thanks{Grants or other notes
%about the article that should go on the front page should be
%placed here. General acknowledgments should be placed at the end of the article.}
}
%\subtitle{Do you have a subtitle?\\ If so, write it here}

\titlerunning{First Magnetic Fields}        % if too long for running head

\author{Lawrence~M.~Widrow,~Dongsu~Ryu, Dominik R. G.~Schleicher,
  \nobreak Kandaswamy~Subramanian, Christos~G. Tsagas and  Rudolf A. ~Treumann
      % Second Author %etc.
}

%\authorrunning{Short form of author list} % if too long for running head

\institute{Lawrence M. Widrow \at
  Department of  Physics, Engineering Physics and Astronomy, Queen's University, Kingston, Ontario, Canada K7L 3N6, \email{widrow@astro.queensu.ca} \\
  \and Dongsu Ryu, \at Department of Astronomy and Space Science,
  Chungnam National University, Daejeon, 305-764,
  Korea\email{ryu@canopus.cnu.ac.kr}
  \\
  \and Dominik Schleicher \at Leiden Observatory, Leiden University,
  P.O.Box 9513, NL-2300 RA Leiden, The Netherlands and
  ESO, Karl-Schwarzschild-Str. 2, D-85748 Garching, Germany \\
  \email{dschleic@eso.org}             \\
  \and Kandaswamy Subramanian \at IUCAA, Post Bag 4, Pune University
  Campus, Ganeshkhind, Pune 411 007, India \\
  \email{kandu@iucaa.ernet.in}
  \\
  \and Christos G. Tsagas \at Department of Physics, Aristotle
  University Thessaloniki, Thessaloniki 54124, Greece\\
  \email{tsagas@astro.auth.gr}
  \\
  \and Rudolf A. Treumann \at ISSI, CH-3012 Bern, Hallerstrasse 6,
  Switzerland, \email{treumann@issibern.ch}
                      %  \\
%             \emph{Present address:} of F. Author  %  if needed
       %   \and
         % S. Author \at
       %      second address
}

\date{Received: date / Accepted: date}
% The correct dates will be entered by the editor

\maketitle

\begin{abstract}

  We review current ideas on the origin of galactic and extragalactic
  magnetic fields.  We begin by summarizing observations of magnetic
  fields at cosmological redshifts and on cosmological scales.  These
  observations translate into constraints on the strength and scale
  magnetic fields must have during the early stages of galaxy
  formation in order to seed the galactic dynamo.  We examine
  mechanisms for the generation of magnetic fields that operate prior
  during inflation and during subsequent phase transitions such as
  electroweak symmetry breaking and the quark-hadron phase transition.
  The implications of strong primordial magnetic fields for the
  reionization epoch as well as the first generation of stars is
  discussed in detail.  The exotic, early-Universe mechanisms are
  contrasted with astrophysical processes that generate fields after
  recombination.  For example, a Biermann-type battery can operate in
  a proto-galaxy during the early stages of structure formation.
  Moreover, magnetic fields in either an early generation of stars or
  active galactic nuclei can be dispersed into the intergalactic
  medium.

% \PACS{PACS code1 \and PACS code2 \and more}
% \subclass{MSC code1 \and MSC code2 \and more}
\end{abstract}

\vspace{0.5cm}
\begin{quote}
{\it There is much to be learned about cosmic magnetic fields. We have
a rather sketchy information about the field distribution on the
largest scales, and the origin of the magnetic fields remains a
mystery.}

\hfill{\small{Alexander Vilenkin 2009}}
\end{quote}

\section{Introduction}
\label{intro}

Magnetic fields are observed in virtually all astrophysical systems,
from planets to galaxy clusters.  This fact is not surprising since
gravitational collapse and gas dynamics, the key processes for
structure formation, also amplify and maintain magnetic fields.
Moreover, the conditions necessary for a magnetic dynamo, namely
differential rotation and turbulence, exist in galaxies which are the
building blocks for large scale structure.  The one notable example
where magnetic fields are searched for but not yet found is in the
surface of last scattering.  All this raises an intriguing question:
When did the first magnetic fields arise?

Numerous authors have suggested that magnetic fields first appeared
in the very early Universe.  There is strong
circumstantial evidence that large scale structure formed from the
amplification of linear density perturbations that originated as
quantum fluctuations during inflation.  It is therefore natural to
consider whether quantum fluctuations in the electromagnetic field
might similarly give rise to large-scale magnetic fields.

Indeed, magnetic fields were almost certainly generated during
inflation, the electroweak phase transition, and the quark-hadron
phase transition but with what strength and on what scale?  And more
to the point, what happened to these fields as the Universe expanded?
Were these early Universe fields the seeds for the magnetic fields
observed in present-day galaxies or clusters?  And even if not, did
they leave an observable imprint on the cosmic microwave background?

Exotic early universe mechanisms for field generation stand in
contrast with mechanisms that operate in the post-recombination
Universe.  There are several ways to generate magnetic fields during
the epoch of structure formation.  At some level, all mechanisms begin
with a battery, a process that treats positive and negative charges
differently and thereby drives currents which in turn generate fields.
A Biermann battery, for example, can (in fact {\it must}) operate
during the formation of a proto-galaxy.  While angular momentum in
proto-galaxies is generated by the tidal torques due to nearby
systems, vorticity arises from gasdynamical processes.  The same
processes also drive currents and hence generate fields, albeit
of small amplitude.

The Biermann battery also operates in compact objects such as
accretion disks and stars.  Since the dynamical timescales for these
systems is relatively short, tiny seed fields are rapidly amplified.  The
magnetic fields in AGN and/or Pop III stars can be expelled into the
proto-galactic medium providing another source of seed fields for
subsequent dynamo action.

In this review, we survey ideas on the generation of magnetic fields.
Our main focus is on mechanisms that operate in the early Universe,
either during inflation, or during the phase transitions that follow.
We contrast these mechanisms with ones that operate during the early
stages of structure formation though we leave the details of those
ideas for the subsequent chapter on magnetic fields and the formation
of large scale structure.  The outline of the chapter is as follows:
In Section 2, we summarize observational evidence for magnetic fields
at cosmological redshifts and on supercluster scales and beyond.  In
Section 3, we discuss the generation of magnetic fields during
inflation.  We make the case that inflation is an attractive arena for
magnetic-field generation but for the fundamental result that
electromagnetic fields in the standard Maxwell theory and in an
expanding, spatially flat, inflating spacetime and massively diluted
by the expansion of the Universe.  However, one can obtain
astrophysically interesting fields in spatially curved metrics or with
non-standard couplings between gravity and electromagnetism.  Section
4 addresses the question of whether fields can be generated during a
post-inflation phase transition.  We will argue that strong fields
almost certainly arise but their scales are limited by the Hubble
radius at these early times.  Only through some dynamical process such
as an inverse cascade (which requires appreciable magnetic helicity)
can one obtain astrophysically interesting fields.  In Section 5 we
take, as given, the existence of strong fields from an early Universe
phase transitions and explore their impact on the post-recombination
Universe.  In particular, we discuss the implications of strong
primordial fields on the first generation of stars and on the
reionization epoch.  Finally, in Section 6 we briefly review
field-generation mechanisms that operate after recombination.  A
summary and some conclusions are presented in Section 7.

\section{Cosmological magnetic fields observed}

The existence of microgauss fields in present-day galaxies and galaxy
clusters is well established.  These fields can be explained by the
amplification of small seed fields over the 13.7\,Gyr history
of the Universe.  However, there is mounting evidence that microgauss
fields existed in galaxies when the Universe was a fraction of its
present age.  Moreover, there are hints that magnetic fields exist
on supercluster scales.  Both of these observations present challenges
for the seed field hypothesis.  In this section, we summarize
observational evidence for magnetic fields at early times and on
cosmological scales and briefly discuss the implications for the
seed field hypothesis.

\subsection{Galactic magnetic fields at intermediate redshifts}

Microgauss fields are found in present-day galaxies of all types as
well as galaxy clusters \citep[see, for example,][]{kronberg94, widrow,
carilli, KZ} .  Perhaps more significant, at least for our purposes, are
observations of magnetic fields in intermediate and high
redshift galaxies.  For example, \cite{kronberg92} found evidence for a
magnetized galaxy at a redshift of $z=0.395$.  To be specific, they
mapped the rotation measure (RM) across the absorption-line quasar,
PKS 1229-021 ($z=1.038$) and determed the residual rotation measure
(RRM -- defined to be the observed roation measure minus the Galactic
rotation measure).  The RRM was then identified with an intervening
galaxy whose magnetic properties were inferred by detailed modelling.

Along similar lines, \cite{bernet} provide a compelling argument for
magnetic fields at $z\sim 1.3$.  Earlier work by \cite{kronberg08}
found a correlation between the {\it spread} in the quasar
RM distribution and their redshift.  The naive expectation is that
the spread in the distribution should decrease with
redshift.  Recall that the polarization angle is proportional to the
square of the wavelength; the proportionality constant is what we
define as the RM.  As electromagnetic radiation propagates from source
to observer, the rotation angle is preserved by the wavelength
increases as $\left (1+z\right )^{-1}$.  Hence, RM is diluted by a
factor $\left (1+z\right )^{-2}$.  In principle, the change in the RM
distribution with redshift could be indicative of a redshift
dependence in quasar magnetic fields.  However, \cite{bernet} sorted
the sample according to the presence of MgII absorption lines and
showed that the RM spread for set of objects with one or more lines
was significantly greater than for the objects with no absorption
lines.  MgII absorption lines arise as the quasar light passes through
the halos of normal galaxies.  The implication is that intervening
galaxies produce both large RMs and MgII absorption lines and hence,
the intervening galaxies must have strong magnetic fields.  Simple
estimates suggest that the fields are comparable to those in
present-day galaxies and that these galaxies are at a redshift of
$z\sim 1.3$.

\cite{athreya} studied 15 radio galaxies with redshifts between
$z\simeq 2$ and $z\simeq 3.13$ and found significant RM's in almost all of them.
Moreover, the RM's were found to differ significantly between the two
radio lobes suggesting that they are intrinsic to the object rather
than due to the Faraday screen of the Galaxy.  The RM's, corrected for
cosmological expansion and with the Galactic contribution removed,
were typically ranged from $100-6000\, {\rm rad\,m^{-2}}$ implying
microgauss field strengths.

\begin{figure}[t!]
\centerline{{\includegraphics[width=0.7\textwidth,height=0.5\textwidth, clip=]{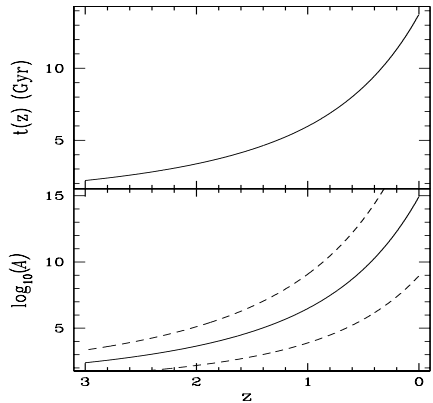} }}
\caption[]{\footnotesize Upper panel: Age of the Universe as a function of redshift
for the cosmology described in the text and in \cite{komatsu}.  Lower
panel: Amplification factor for a seed magnetic field assuming
exponential growth with one of three choices for the growth rate: $\Gamma = 1.5 \,{\rm Gyr}^{-1}$
$\Gamma = 2.5 \,{\rm Gyr}^{-1}$ or $\Gamma = 1.5 \,{\rm Gyr}^{-1}$}
{\footnotesize }\label{ageofU}
\vspace{-0.3cm}
\end{figure}

Observations of magnetic fields at intermediate redshift imply a
shorter time over which the dynamo can operate.  Consider the standard
$\Lambda$CDM cosmological model with $\Omega_m = 0.272$ and
$\Omega_\Lambda = 0.728$ where $\Omega_m$ and $\Omega_\Lambda$ are the
density, in units of the critical density, for matter (both baryons and
dark matter) and dark energy \cite{komatsu}.  In Figure \ref{ageofU},
we show the age of the Universe as a function of redshift for this
cosmology.  We also show the amplication factor for a seed magnetic
field where we assume exponential growth and one of three choices for the
growth rate, $\Gamma = 1.5 \,{\rm Gyr}^{-1}, ~2.5\, {\rm Gyr}^{-1},$ or $~3.5\,
{\rm Gyr}^{-1}$.  A seed field of only $10^{-21}\,{\rm G}$ is required
to reach microgauss strength assuming $\Gamma \,2.5{\rm Gyr}^{-1}$.
However, for the same growth rate, a $10^{-11}\,{\rm G}$ seed field is
required to reach microguass strengths by a redshift $z=1.3$

\subsection{Magnetic fields on supercluster scales and beyond}

\cite{kim89} used the Westerbork Synthesis Radio Telescope to map the
Coma cluster and its environs at 326 MHz.  Their results provide what
remains the best direct evidence for magnetic fields on supercluster
scales.

Recently, \cite{NV} argued that the deficit of GeV gamma rays in the
direction of TeV gamma-ray sourses yields a {\it lower bound} of
$3\times 10^{-16}\,{\rm G}$ on the strength of intergalactic magnetic
fields.  The reasoning goes as follows: TeV gamma rays and photons
from the diffuse extragalactic background light produce $e^\pm$ pairs
which, in turn, inverse Compton scatter off photons from the cosmic
microwave background (CMB).  The scattered CMB photons typically have
energies in the GeV range.  In the absence of appreciable magnetic
fields, these secondary photons contribute to the overall emission
toward the original TeV source.  However, magnetic fields will deflect
the intermediate $e^\pm$ pairs.  Comparison of model predictions with
the observed spectrum from HESS Cherenkov Telescopes and upper limits
from the NASA Fermi Gamma-Ray telescope hint at just such a deficit
which \cite{NV} use to derive their lower limit on the magnetic
fields.

\section{Magnetic fields from inflation}\label{ssEMFRC}
%%%%%%%%%%%%%%%%%%%%%%%%%

\subsection{General considerations}

The hierarchical clustering scenario provides a compelling picture for
the formation of large-scale structure.  Linear density perturbations
from the early Universe grow via gravitational instability.
Small-scale objects form first and merge to create systems of
ever-increasing size.  The spectrum of the primordial density
perturbations, as inferred from the CMB anisotropy spectrum and
various statistical measures of large scale structure (e.g., the
galaxy two-point correlation function) is generally thought to be
scale-invariant and very close to the form proposed by
\cite{zeldovich70}.  One of the great successes of inflation is that
it leads to just such a spectrum \citep{guthpi,hawking,starobinskii}.
It is therefore natural to ask whether a similar mechanism might
generate large-scale magnetic fields.

\begin{figure}[t!]
\centerline{{\includegraphics[width=0.7\textwidth,height=0.5\textwidth, clip=]{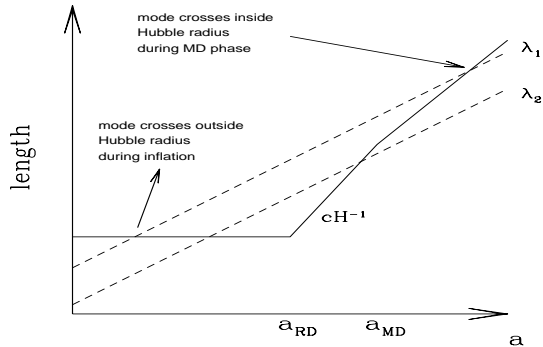} }}
\caption[]{\footnotesize Evolution of the physical size for the Hubble
  radius (solid curve) and two scales, $\lambda_1$ and $\lambda_2$
  (dashed curves) as a function of scale factor $a$.  Shown is the
  point at which the scale $\lambda_1$ crosses outside the Hubble
  radius during inflation and back inside the Hubble radius during the
  matter-dominated phase.  The scale factors $a_{RD}$ and $a_{MD}$
  correspond to the start of the radiation and matter dominated
  epochs, respectively.  Here, $\lambda_1$ enters the Hubble volume
  during the matter-dominated epoch while $\lambda_2$ enters the
  Hubble volume during the radiation-dominated epoch.  }
\label{inflation}
\vspace{-0.3cm}
\end{figure}

In order to understand the meaning and significance of the results for
density perturbations, we must say a few words about horizons in
cosmology.  The Hubble radius, essentially, the speed of light divided
by the Hubble parameter, sets the maximum scale over which
microphysical processes can operate.  In a radiation or
matter-dominated Universe the Hubble scale is equal, up to constants
of order unity, to the causal scale which is defined as the distance
over which a photon could have propagated since the Big Bang.
Moreover, the Hubble scale grows linearly with time $t$ in a
radiation or matter-dominated Universe whereas the physical size of an
object associated with a fixed comoving (or present-day) scale grows
as $t^2$ or $t^{3/2}$.  Therefore, a physical scale crosses inside the
Hubble radius with smaller objects crossing earlier than larger ones.
The point is illustrated in Figure \ref{inflation} and explains why it
is so difficult to generate {\it large-scale} magnetic fields in the
early Universe but after inflation.

During inflation, the Hubble parameter is approximately constant (the
spacetime is approximately de Sitter) and a physical scale that is
initially inside the Hubble radius will cross outside the Hubble
radius or Hubble scale, at least until some time later in the history of the Universe
(Figure \ref{inflation}).  We therefore have the potential for
microphysical processes to operate on large scales during inflation
with the consequences of these processes becoming evident much later when the
scale re-enters the Hubble radius.

Inflation provides the dynamical means for generating density
perturbations: quantum mechanical fluctuations in the de Sitter
phase excite modes on the Hubble scale with an energy density
set by the Hubble parameter.  Therefore, to the extent that the Hubble
parameter is constant during inflation, the energy density in modes
as they crosses outside the Hubble radius will be scale-independent.

There is one further and crucial part to the story.  During inflation,
the energy density of the Universe is approximately constant.  It is,
indeed, the constant energy density that drives the exponential
expansion of the de Sitter phase.  Naively, we expect that the
energy density in some (relativistic) fluctuation scales as
$a^{-4}$ and is therefore diluted by an enormous factor.  However, for
density perturbations with a coherence length greater than the Hubble radius, the
energy density scales as $a^{-2}$.  The ratio, $r$, of the energy
density in the fluctuation relative to the background density
therefore decreases as $a^{-2}$ during inflation but {\it grows} as
$a^2$ or $a$ during the subsequent radiation and matter dominated
phases.  The net result is that $r$ is the same when the fluctuation
re-enters the Hubble radius as when it crossed outside the Hubble
radius during inflation.  The $a^{-2}$ behaviour is often referred to
as super-adiabatic growth since it implies that the energy density in
the fluctuation grows as $a^2$ relative to the energy density in the
radiation field.

The equation of motion for a scalar field, $\phi$, in a curved
spacetime is
\begin{equation}
\nabla^\alpha\nabla_\alpha\phi - \xi R\phi - \frac{dV}{d\phi} = 0
\end{equation}
where $R$ is the Ricci scalar, $V$ is the scalar field potential (the
driving term for inflation, but ignored in the present discussion),
and $\xi$ is a dimensionless constant.  Super-adiabatic growth occurs
for a minimally-coupled scalar field ($\xi=0$).  On the other hand,
the field evolves adiabatically (energy density scales in the same way
as radiation) if it is conformally coupled ($\xi=1/6$).

The equation of motion for energy density perturbations is identical
to that for the minimally coupled scalar field.  On the other hand,
electromagnetism is conformally-coupled to gravity (at least in the
simplest version of the theory) and any de Sitter-induced quantum
fluctuations will have an energy density that scales as $a^{-4}$.  One
finds that the relative strength a magnetic mode at the end of
inflation is
\begin{equation} {\rho_B\over\rho_t}\simeq 10^{-78}\left({M\over
      m_{Pl}}\right)^4 \left({M\over10^{14}\,{\rm GeV}}\right)^{-8/3}
  \left({T_{RH}\over10^{10}\,{\rm GeV}}\right)^{-4/3}\lambda^{-4}\,, \label{adr1}
\end{equation}
where $\lambda$ is the comoving scale of the mode and $m_{Pl} \simeq
10^{19}\,{\rm GeV}$ is the Planck mass.  The above ratio also depends
on the energy scale of our inflation model ($M$) and on the associated
reheating temperature ($T_{RH}$).  Assuming efficient reheating
(i.e.~setting $T_{RH}\simeq M$), expression (\ref{adr1}) yields
\begin{equation}
{\rho_B\over\rho_t}\simeq 10^{-104}\lambda^{-4}\,.  \label{adr2}
\end{equation}

During the radiation era, the energy density of the universe is
dominated by that of the relativistic species
(i.e.~$\rho_t=\rho_{\gamma}\propto a^{-4}$) and the high conductivity
of the matter has been restored.  As a result, the magnetic flux is
conserved (i.e.~$B\propto a^{-2}$) and the dimensionless ratio
\begin{equation}
r\equiv {\rho_B\over\rho_{\gamma}}\simeq 10^{-104}\lambda^{-4}\,,  \label{adr3}
\end{equation}
remains constant.  This result implies a field strength no greater
than $10^{-50}\,{\rm G}$ on a comoving scales of order $10\,{\rm kpc}$
which are the scales relevant for the galactic dynamo.

We are lead to the conclusion that inflation-produced magnetic fields
are astrophysically uninteresting.  However, this `negative' result
holds for the standard formulation of Maxwell's equations and under
the assumption of a spatially-flat FLRW cosmology.  In the next
sections, we show that super-adiabatic growth can occur in various
``open'' cosmologies and in models in which certain additional
couplings between electromagnetic field and gravity are included.

\subsection{Maxwell's equations}\label{sssMEs}
%%%%%%%%%%%%%%%%%%%%%%%%%%%%%%%%%%%%%%%%%%%%%%%%%

The Maxwell field may be invariantly described by the
antisymmetric Faraday tensor, $F_{ab}$.  Relative to an observer
moving with 4-velocity $u_a$, we can write $F_{ab}= 2u_{[a}E_{b]}+
\varepsilon_{abc}B^c$, where $E_a=F_{ab}u^b$ and
$B_a=\varepsilon_{abc}F^{bc}/2$ represent the electric and magnetic
components of the EM field respectively. Maxwell's equations \textbf{then} split into two pairs of propagation and constraint
equations~\citep{T1,BMT}. The former consists of
\begin{equation}
\dot{E}_{\langle a\rangle}= -{2\over3}\,\Theta E_a+ \left(\sigma_{ab}+\omega_{ab}\right)E^b+ \varepsilon_{abc}A^bB^c+ {\rm curl}{B}_a- \mathcal{J}_a\,,  \label{M1}
\end{equation}
and
\begin{equation}
\dot{B}_{\langle a\rangle}= -{2\over3}\,\Theta B_a+ \left(\sigma_{ab}+\omega_{ab}\right)B^b- \varepsilon_{abc}A^bE^c- {\rm curl}{E}_a\,,  \label{M2}
\end{equation}
which may be seen as the 1+3 covariant analogues of the Amp\`{e}re and the
Faraday laws respectively.  The constraints, on the other hand, read
\begin{equation}
{\rm D}^aE_a= \mu- 2\omega^aB_a \hspace{15mm} {\rm and} \hspace{15mm}
{\rm D}^aB_a= 2\omega^aE_a\,,  \label{M34}
\end{equation}
providing the 1+3 forms of Coulomb's and Gauss' laws
respectively. In the above $\Theta$, $\sigma_{ab}$, $\omega_{ab}$ and $A_a$ respectively represent the volume expansion, the shear, the vorticity and the 4-acceleration associated with the observer's motion \citep{tsagas2008}. Also, $\mathcal{J}_a$ and $\mu$ are the 3-current and the charge densities respectively.

%Maxwell's formulae also lead to the wave equation for the
%electromagnetic tensor.  In particular, applying the Ricci identities
%to the Faraday tensor, one arrives at~\cite{N}
%\begin{equation}
%\nabla^2F_{ab}= -2R_{abcd}F^{cd}- 2R_{[a}{}^cF_{b]c}- 2\nabla_{[a}J_{b]}\,,  \label{N2Fab}
%\end{equation}
%where $\nabla^2=\nabla^a\nabla_a$ is the d'Alembertian and $R_{abcd}$,
%$R_{ab}$ are the Riemann and Ricci tensors respectively. The
%geometrical terms in the right-hand side of (\ref{N2Fab}) reflect the
%vector nature of the electromagnetic components and the geometrical
%interpretation of gravity that general relativity advocates. The two
%guarantee that the Maxwell field is, as yet, the only known energy
%source that couples directly to gravity through both the Einstein
%equations and the Ricci identities.

These equations, together with
the Einstein equations and the Ricci identities, lead to wave
equations for the electric and magnetic fields.   For example,
linearised on a Friedmann-Lema\^{i}tre-Robertson-Walker (FLRW) background, the wave equations of the electric and the magnetic components of the Maxwell field read~\citep{T1}
\begin{equation}
\ddot{E}_a- {\rm D}^2E_a= -5H\dot{E}_a+ {1\over3}\,\kappa(\rho+3p)E_a- 4H^2E_a- {1\over3}\,\mathcal{R}E_a- \dot{\mathcal{J}}_a- 3H\mathcal{J}_a  \label{ddotEa}
\end{equation}
and
\begin{equation}
\ddot{B}_a- {\rm D}^2B_a= -5H\dot{B}_a+ {1\over3}\,\kappa(\rho+3p)B_a- 4H^2B_a- {1\over3}\,\mathcal{R}B_a+ {\rm curl}{\mathcal{J}}_a\,,  \label{ddotBa}
\end{equation}
respectively. Note that $H=\dot{a}/a=\Theta/3$ is the background Hubble parameter, $\mathcal{R}=6K/a^2$ represents the Ricci scalar of the
spatial sections (with $K=0,\pm1$ being the associated 3-curvature
index), while $\kappa=8\pi G$ is the gravitational constant.  The
3-Ricci term arises from the purely geometrical coupling
between the electromagnetic field and the spacetime curvature.
\footnote{For the full expressions
  of Eqs.~(\ref{ddotEa}) and (\ref{ddotBa}), written in a general
  spacetime, the reader is referred to~\cite{T1}. There, one can also
  see how the different parts of the geometry (i.e.~the Ricci and the
  Weyl fields) can affect the propagation of electromagnetic signals.}
We will return to this particular interaction to examine the way it
can affect the evolution of cosmological magnetic fields.

\subsection{Ohm's law in the expanding Universe}\label{sMEFRWBs1}

The literature contains various expressions of Ohm's law, which
provides the propagation equation of the electric 3-current. For a
single fluid at the limit of resistive magnetohydrodynamics (MHD),
Ohm's law takes the simple form~\citep{G,J}
\begin{equation}
\mathcal{J}_a= \varsigma E_a\,,  \label{Ohm}
\end{equation}
with $\varsigma$ representing the electric conductivity of the
matter. In highly conducting environments,
$\varsigma\rightarrow\infty$ and the electric field vanishes. This is
the familiar ideal-MHD approximation where the electric currents keep
the magnetic field frozen-in with the charged fluid.
Conversely, when the conductivity is very low,
$\varsigma\rightarrow0$. Then, the 3-currents vanish despite the
presence of nonzero electric fields. Here, we will consider these two
limiting cases. For any intermediate case, one needs a model for the
electrical conductivity of the cosmic medium.

\subsection{Adiabatic decay of magnetic fields in a spatially flat FLRW cosmology}\label{sMEFRWBs2}

Consider the case of a poorly conductive environments where there are
no 3-currents.  The wave equation, (\ref{ddotBa}), then reduces to \citep{T1}
\begin{equation}
\ddot{B}_a- {\rm D}^2B_a= -5H\dot{B}_a+ {1\over3}(\rho+3p)B_a- 4H^2B_a- \mathcal{R}_{ab}B^b\,.  \label{lddBa}
\end{equation}
To simplify the above, we introduce the rescaled the magnetic field
$\mathcal{B}_a=a^2B_a$ and employ conformal time, $\eta$, where
$\dot{\eta}=1/a$. Then, on using the harmonic splitting
$\mathcal{B}_a= \sum_n\mathcal{B}_{(n)}\mathcal{Q}_a^{(n)}$ -- so that
${\rm D}_a\mathcal{B}_{(n)}=0$, expression (\ref{lddBa}) takes the
compact form
%\footnote{We use vector harmonics, with
% $\dot{\mathcal{Q}}_a^{(n)}=0={\rm D}^a\mathcal{Q}_a^{(n)}$ and ${\rm
%  D}^2\mathcal{Q}_a^{(n)}=-(n/a)^2\mathcal{Q}_a^{(n)}$. The
%eigenvalues depend on the background curvature. Thus, $n$ is
%continuous, with $n^2\geq0$, when $K=0,-1$ and discrete, with
%$n^2\geq3$, if $K=+1$.}
\begin{equation}
\mathcal{B}_{(n)}^{\prime\prime}+ n^2\mathcal{B}_{(n)}= -2K\mathcal{B}_{(n)}\,,  \label{cB''}
\end{equation}
with the primes denoting conformal-time derivatives and
$K=0,\pm1$ \citep{T1}. Note the magneto-curvature term in the right-hand side of (\ref{cB''}), which shows that the magnetic
evolution also depends on the spatial geometry of the FLRW spacetime.

When the background has Euclidean spatial hypersurfaces, the
3-curvature index is zero (i.e.~$K=0$) and expression (\ref{cB''})
assumes the Minkowski-like form
\begin{equation}
\mathcal{B}_{(n)}^{\prime\prime}+ n^2\mathcal{B}_{(n)}=0\,.  \label{EcB''}
\end{equation}
This equation accepts the oscillatory solution $\mathcal{B}_{(n)}= C_1\sin(n\eta)+C_2\sin(n\eta)$, which recasts into
\begin{equation}
B_{(n)}= \left[\mathcal{C}_1\sin(n\eta) +\mathcal{C}_2\cos(n\eta)\right]\left({a_0\over a}\right)^2\,,  \label{adB}
\end{equation}
for the actual $B$-field. In other words, the adiabatic
($B_{(n)}\propto a^{-2}$) depletion of the magnetic component is
guaranteed, provided the background spacetime is a spatially flat and
the electrical conductivity remains very poor.  This result leads
directly to Equation \ref{adr1}.

The adiabatic decay-law also holds in highly conductive
environments. There, $\varsigma\rightarrow\infty$ and, according to
Ohm's law (see Eq.~(\ref{Ohm}) the electric field vanishes in the
frame of the fluid. As a result, Faraday's law (see Eq.~(\ref{M2}))
linearises to
\begin{equation}
\dot{B}_a= -2HB_a\,,  \label{MHDadB}
\end{equation}
around an FLRW background. The above ensures that $B_a\propto a^{-2}$
on all scales, regardless of the equation of state of the matter and
of the background 3-curvature.  This result leads directly to Equation \ref{adr3}.

%%%%%%%%%%%%%%%%%%%%%%%%%%%%%%%%%%%%%%%%%%%%%%%%%%%%%%%%%%%%%%%

\subsection{Superadiabatic \textbf{magnetic} amplification in spatially open FLRW models}\label{ssAMDFRWMs}

The ``negative'' results discussed at the beginning of this section
have been largely attributed to the conformal invariance of Maxwell's
equations and to the conformal flatness of the Friedmannian
spacetimes. The two are thought to guarantee an adiabatic decay-rate
for all large-scale magnetic fields at all times. Solution
(\ref{adB}), however, only holds for the spatially flat FLRW
cosmology.  Although all three FLRW universes are
conformally flat, only the spatially flat model is
globally conformal to Minkowski space. For the rest, the conformal
mappings are local.
%\footnote{Based on that, changes in the magnetic
% evolution should only occur on large scales, where the curvature
% effects are prominent and the conformal mapping between the FLRW
% spacetimes and the Minkowski space breaks down. This is indeed what
% happens (see Eq.~(\ref{pm1cB''})).}
Put another way,
in spatially curved Friedmann universes, the
conformal factor is no longer the cosmological scale factor but has an
additional spatial dependence~\citep{S,KB}.  The
wave equation of the rescaled magnetic field ($\mathcal{B}_a=a^2B_a$)
takes the simple Minkowski-like form (\ref{EcB''}) only on FLRW
backgrounds with zero 3-curvature. In any other case, there is an
additional curvature-related term (see expressions (\ref{lddBa}) and
(\ref{cB''})), which reflects the non-Euclidean spatial geometry of
the host spacetime. As a result, when linearised around an FLRW
background with nonzero spatial curvature, the magnetic wave equation
reads~\citep{TK,BT}
\begin{equation}
\mathcal{B}_{(n)}^{\prime\prime}+ \left(n^2\pm2\right)\mathcal{B}_{(n)}= 0\,,  \label{pm1cB''}
\end{equation}
with the plus and the minus signs indicating the spatially closed and
the spatially open model respectively. Recall that in the former case
the eigenvalues are discrete (with $n^2\geq3$), while in the latter
continuous (with $n^2\geq0$). As expected, in either case, the
curvature-related effects fade away as we move on to successively
smaller scales (i.e.~for $n^2\gg2$).

%%%%%%%%%%%%%%%%%%%%%%%%%

Following (\ref{pm1cB''}), on FLRW backgrounds with spherical spatial
hypersurfaces, the $B$-field still decays adiabatically. The picture
changes when the background FLRW model is open. There, the hyperbolic
geometry of the 3-D hypersurfaces alters the nature of the magnetic
wave equation on large enough scales (i.e.~when $0<n^2<2$). These
wavelengths include what we may regard as the largest subcurvature
modes (i.e.~those with $1\leq n^2<2$) and the supercurvature lengths
(having $0<n^2<1$). Note that eigenvalues with $n^2=1$ correspond to
the curvature scale with physical wavelength
$\lambda=\lambda_K=a$~\citep{LW}. Here, we will focus on the largest
subcurvature modes.
%\footnote{Although they often omitted
%supercurvature modes are necessary if we want perturbations with
%correlations lengths larger than the curvature radius \citep[see][]{LW}
%for further discussion).}

In line with~\cite{TK} and \cite{BT}, we introduce the scale-parameter
$k^2=2-n^2$, with $0<k^2<2$. Then, $k^2=1$ indicates the curvature
scale, the range $0<k^2<1$ corresponds to the largest subcurvature
modes and their supercurvature counterparts are contained with in the
$1<k^2<2$ interval. In the new notation and with $K=-1$,
Eq.~(\ref{pm1cB''}) reads
\begin{equation}
\mathcal{B}_{(n)}^{\prime\prime}- k^2\mathcal{B}_{(n)}= 0\,,  \label{-1cB''}
\end{equation}
with the solution given by $\mathcal{B}_{(k)}=C_1\sinh(|k|\eta)+ C_2\cosh(|k|\eta)$. Written in terms of the actual magnetic field, the latter takes the form
\begin{equation}
B_{(k)}= \left[\mathcal{C}_1{\rm e}^{|k|(\eta-\eta_0)}+ \mathcal{C}_2{\rm e}^{-|k|(\eta-\eta_0)}\right] \left({a\over a_0}\right)^{-2}\,.  \label{-1B1}
\end{equation}
Magnetic fields that obey the above can experience superadiabatic
amplification without modifying conventional electromagnetism and
despite the conformal flatness of the FLRW
host.
%\footnote{Superadiabatic amplification does not imply
%amplification per se, but decay at a slower than the adiabatic
%pace. The concept was originally introduced in gravitational-wave
%studies.}
For instance, during the radiation era, the scale factor
of an open FLRW universe evolves as $a\propto\sinh(\eta)$. Focusing on
the curvature length, for simplicity, we may set $|k|=1$ in
Eq.~(\ref{-1B1}). On that scale, the dominant magnetic mode never
decays faster than $B_{(1)}\propto a^{-1}$. The $B$-field has been
superadiabatically amplified.\footnote{Mathematically, the most
  straightforward case of superadiabatic amplification occurs on a
  Milne background. The latter corresponds to an empty spacetime with
  hyperbolic spatial geometry and can be used to describe a low
  density open universe. The scale factor of the Milne universe
  evolves as $a\propto{\rm e}^{\eta}$, which substituted into solution
  (\ref{-1B1}) leads to
\begin{equation}
B_{(k)}= \mathcal{C}_5\left({a_0\over a}\right)^{|k|-2}+ \mathcal{C}_6\left({a_0\over a}\right)^{-|k|-2}\,.  \label{MilneB}
\end{equation}
Consequently, all magnetic modes spanning scales with $0<k^2<2$ are
superadiabatically amplified. Close to the curvature scale, that is
for $k^2\rightarrow1$, the dominant magnetic mode is $B_{(1)}\propto
a^{-1}$. Stronger amplification is achieved on supercurvature lengths,
with $B_{(k)}\propto a^{\sqrt{2}-2}$ at the homogeneous limit (i.e.~as
$k^2\rightarrow2$).}

Analogous amplification also occurs in open FLRW universes with an
inflationary (i.e.~$p=-\rho$) equation of state. There, the scale
factor evolves as \citep{tsagas2008}
\begin{equation}
a= a_0\left({1-{\rm e}^{2\eta_0}\over1-{\rm e}^{2\eta}}\right) {\rm e}^{\eta-\eta_0}\,,  \label{-1infFRW}
\end{equation}
where $\eta,\eta_0<0$. Substituting the above into Eq.~(\ref{-1B1}), we find that near the curvature scale (i.e.~for $|k|\rightarrow1$) the magnetic evolution is given by
\begin{equation}
B_{(1)}= \mathcal{C}_3\left(1-{\rm e}^{2\eta}\right)\left({a\over a_0}\right)+ \mathcal{C}_4{\rm e}^{-\eta}\left({a_0\over a}\right)^2\,,  \label{-1B2}
\end{equation}
with $\mathcal{C}_3$, $\mathcal{C}_4$ depending on the initial conditions. This result also implies a superadiabatic type of amplification for the $B$-field, since the dominant magnetic mode never decays faster than $B_{(1)}\propto a^{-1}$. The adiabatic decay rate is only recovered at the end of inflation, as $\eta\rightarrow0$.\footnote{The magnetogeometrical interaction and the resulting effects are possible because, when applied to spatially curved FLRW models, inflation does not lead to a globally flat de Sitter space. Although the inflationary expansion dramatically increases the curvature radius of the universe, it does not change its spatial geometry. Unless the universe was perfectly flat from the beginning, there is always a scale where the 3-curvature effects are important. It is on these lengths that primordial $B$-fields can be superadiabatically amplified.}

The strength of the residual magnetic field is calculated in a way
analogous to that given in the previous section. In particular, when
$B\propto a^{-1}$, we find that
\begin{equation}
r= {\rho_B\over\rho_{\gamma}}\simeq 10^{-51} \left({M\over10^{17}}\right)^{8/3} \left({T_{RH}\over10^9}\right)^{-2/3}\lambda^{-2}\,,  \label{SAr}
\end{equation}
by the end of inflation~\citep{TK,BT}. As in Eq.~(\ref{adr1}), $M$ and
$T_{RH}$ are measured in GeV, while $\lambda$ is given in Mpc. Thus,
the higher the scale of inflation, the lower the reheatint temperature
and also the smaller the number of e-folds, the stronger the
amplification. Setting $M\sim10^{17}$~GeV and $T_{RH}\sim10^9$~GeV,
for example, the current (comoving) magnetic strength varies between
$\sim10^{-35}$ and $\sim10^{-33}$~Gauss on scales close to the present
curvature scale. The latter lies between $\sim10^4$ and $\sim10^5$~Mpc
when $1-\Omega\sim10^{-2}$ today.  We note that the condition
$1-\Omega\sim10^{-2}$ is within the values allowed by recent analysis
of the WMAP observations \citep{komatsu}.  These lengths are far larger than
$10$~kpc; the minimum magnetic size required for the dynamo to
work. Nevertheless, once the galaxy formation starts, the field lines
should break up and reconnect on scales similar to that of the
collapsing protogalactic cloud. Note that the above quoted strengths
assume that the ratio $r=\rho_B/\rho_{\gamma}$ remains constant after
inflation. As we have seen earlier, however, magnetic fields spanning
lengths close to the curvature scale are superadiabatically amplified
during the radiation era as well. When this is also taken into
account, the residual $B$-field increases further and can reach
magnitudes of up to $10^{-16}$~G at present.

Galactic-scale magnetic fields of strength $10^{-16}$ are stronger
than those generated by many of the other scenarios considered in the
literature (see below) and may be strong enough to seed the galactic
dynamo.  Recall, however, that inflation was introduced to avoid
various shortcomings of the standard cosmological model including the
so-called flatness problem.  Essentially, inflation is meant to
inflate away (push to extremely large scales) any curvature that might
exist in the pre-inflationary Universe.  Evidently, for superadiabatic
growth of magnetic field without explicit conformal symmetry breaking
(as described in the next section) one requires enough inflation to
push the curvature scale beyond our present Hubble volume, but not too
far beyond.  To be quantitative, the most recent analysis by the WMAP
group \citep{komatsu} finds that the effective energy density
parameter for curvature, $\Omega_k \equiv 1 - \Omega_\Lambda -
\Omega_m$, is constrained to be $-0.0133 < \Omega_k < 0.0084$.  Should
future measurements find $\Omega_k$ to be inconsistent with zero {\it
  and positive}, they would lend credence to the idea of
superadiabatic magnetic amplification by the mechanism described above.

% Magnetic fields with the above quoted strengths are far stronger
% than any of their conventionally produced counterparts. Such
% magnitudes are usually achieved outside classical electromagnetism,
% or beyond the standard model (see~\cite{TW} and~\cite{R, CKM, BOT,
%   K, LP, CC} for a representative though incomplete list).  Although
% fields of this strength cannot affect primordial nucleosynthesis or
% the CMB spectrum, they could provide the seed galactic dynamo
% requirements and are therefore of astrophysical interest. Finally,
% we should also note very recent reports indicating the presence of
% coherent magnetic fields in the intergalactic space with strengths
% intriguingly close to those quoted here~\cite{AK,NV}.

\subsection{Inflation-produced magnetic fields via non-conformal couplings}

As pointed out above, electromagnetism is conformally coupled to
gravity and therefore, in a spatially flat, FLRW cosmology, the
magnetic fields generated during inflation decay adiabatically and are
therefore of negligible astrophysical importance.  \cite{TW} pointed
out that by adding additional terms to the Lagrangian such as $RA^2$
and $R_{\mu\nu}A^\mu A^{\nu}$ one explicitly breaks conformal
invariance and can essentially force the magnetic field to behave like
a minimally-coupled scalar field.

The terms introduced by \cite{TW} also break gauge invariance and
hence induce an effective mass for the photon whose size depends on
the spacetime curvature.  In fact, the effective mass-squared is
negative for the case where the magnetic field behaves as a
minimally-coupled scalar field.  The negative mass-squared signals an
instability in the semi-classical equations for the field and can be
viewed as the origin of the superadiabaticity.  These terms also give
rise to ghosts in the theory which signal in instability of the vacuum
\citep{hcp1,hcp2}.  A theory with ghosts is internally consistent only
as an effective low-energy theory and hence is only valid below some
energy scale, $\Lambda$.  \cite{hcp2} argue that $\Lambda\simle {\rm
  MeV}$ but this scale is well below the scales assumed in models for
inflation-produced magnetic fields.  Thus, their results call into
question the viability of the mechanism.

Numerous authors have attempted to find more effective and more
natural ways to break conformal invariance.  Indeed, \cite{TW},
recognizing the potential difficulties associated with the
$RA^2$-terms considered terms such as $RF^2$ which break conformal
invariance but not gauge invariance.  \cite{R} demonstrated that
appreciable magnetic fields couple be produced during inflation if the
electromagnetic field couples to the inflaton field, $\Phi$, through a
term of the form $e^{\alpha\Phi}F_{\mu\nu}F^{\mu\nu}$ where $\alpha$
is a constant.  In his model, the inflaton potential is also an
exponential: $V\left (\Phi\right )\propto \exp{\left (-q\Phi\right
  )}$.  An attractive feature of this model is that is preserves gauge
invariance since the additional term is constructed from the Maxwell
tensor, $F$, rather than the gauge field, $A$.  Along similar lines
\cite{GGV} demonstrated that magnetic fields of sufficient strength to
seed the galactic dynamo could be produced in a string-inspired
cosmology.  In their model, the electromagnetic field is coupled to
the dilaton which, in turn, is coupled to gravity.  The dilaton is a
scalar field that naturally arises in theories with extra dimensions
and whose vacuum expectation value effectively controls Newton's
constant.  A detailed and critical analysis of magnetic-field
production in string-inspired models of inflation can be found in
\cite{MY} \citep[see, also, the review by][]{KS}.  There, particular
attention is paid to a potential back-reaction problem which arises if
the {\it electric} fields produced along with the magnetic fields
have an energy density comparable to the background energy density.

\section{Magnetic fields from early universe phase transitions}

\begin{figure}[t!]
\centerline{{\includegraphics[width=0.7\textwidth,height=0.5\textwidth, clip=]{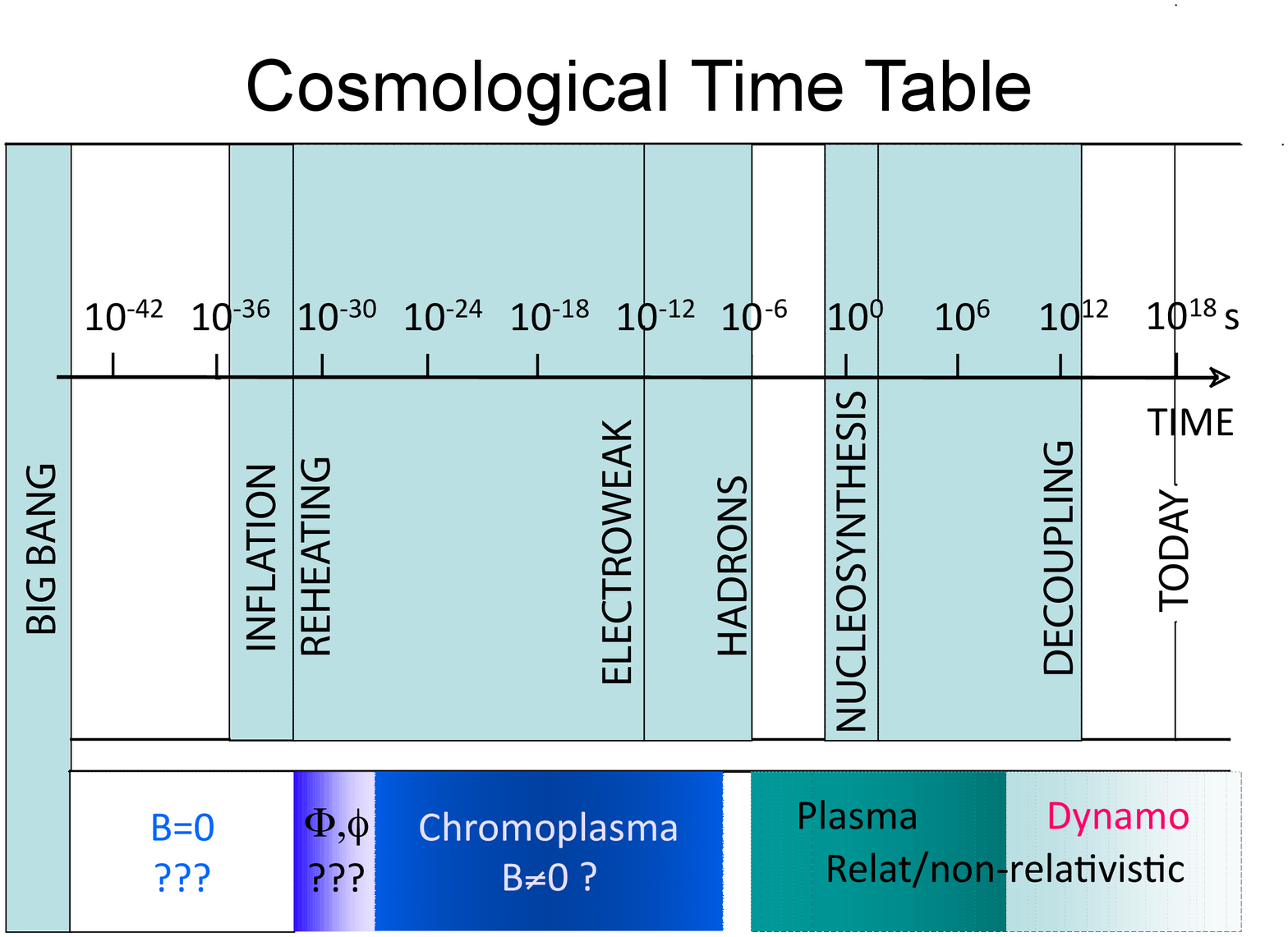} }}
\caption[ ] {\footnotesize The cosmological periods. Starting with
reheating the quantum chromodynamic quark-gluon-plasma period lasts
until hadron formation. Non-Abelian Weibel instabiiities may be
responsible for generating seed magnetic fields during this time which
in presence of free energy grow from non-Abelian thermal
fluctuations. At later times the period of relativistic and classical
plasma sets on , Here seed magnetic fields can be generated from
thermal background electrodynamic fluctutations if thermal
anisotropies or beams can exist in this regime.}\label{rt-weibel-1}
\vspace{-0.3cm}
\end{figure}

The early universe was characterized by a series of phase transitions
in which the nature of particles and fields changed in fundamental
ways (see, Figure \ref{rt-weibel-1}).  For example, electroweak
symmetry breaking marked the transition from a high-energy regime in
which the $W$ and $Z$ bosons and the photon were effectively massless
and interchangeable to one in which the $W$ and $Z$ bosons were heavy
while the photon remained massless.  The transition also marked the
emergence of two distinct forces: electromagnetism and the weak nuclear
force.  Likewise, the Quark-Hadron phase transition marked the
transition from the free quark-gluon phase to one in which quarks were
locked into baryons.  Both of these transitions have the potential to
generate strong magnetic fields since they involve the release of an
enormous amount of free energy and since they involve charged
particles which can, in turn, drive currents.  Indeed, strong magnetic
fields are almost certainly generated.  The question is one of
physical scale since the Hubble radius is so small at these early
times.

\subsection{General considerations}

At any phase in the history of the Universe, the strength of the
fields generated by some microphysical process is limited by
equipartition with the background energy density.
Moreover, the maximum scale for magnetic fields generated by a
microphysical process is set by the Hubble radius.  For a
radiation-dominated Universe, the energy density of the Universe is
given by $\rho \propto g_*T^4$ where $T$ is the temperature of the thermal
bath and $g_*$ counts the effective number of relativistic degrees of
freedom \citep[see, for example,][]{KT, P}.  The Hubble radius
is given by $L_H = c/H$ where $H=a^{-1} da/dt \propto T^2$ is the
Hubble parameter.  Numerically, we have
\begin{equation}
B_{\rm max}\left (l = H^{-1},\,T\right ) = B_{\rm equipartition} \left (T\right )
\simeq 10^{18} \,{\rm Gauss}\,\left (\frac{T}{150\,{\rm MeV}}\right )^2
\end{equation}
for
\begin{equation}
l ~\simeq ~ 100\,{\rm cm}\left (\frac{150\,{\rm MeV}}{T}\right )^2~.
\end{equation}
In practive, the strength of fields generated during some early
Universe phase transition will be well below the value set by
equipartition and have a length scale significantly smaller than the
Hubble radius.  We therefore write $B(l,T) = fB_{\rm max}$ for
$l=gc/H$ and $f$ and $g$ are constants.

We are typically interested in fields on scales much larger than the
Hubble scale of some early Universe phase transition.  \cite{hogan}
argued on purely geometric grounds, that the field strength on a large
scale $L$ due to small-scale cells of size $l$ with field strength
$B(l)$ will be $B(L) = B(l)\left (l/L\right )^{3/2}$.  Moreover, in
the absence of some dynamical effect, the field strength will be
diluted by the expansion as $a^{-2}$.  Thus, a field $B(l,T)$
generated prior to recombination will lead to a field at recombination
with strength
\begin{equation}
 B\left (L,\,T_{\rm rec}\right ) ~ = ~ B\left (l,\,T\right )
\left (\frac{T_{\rm rec}}{T}\right )^2\,\left (\frac{l}{L}\right )^{3/2}~.
\end{equation}
More generally we have the following scaling:
\begin{equation}
B\left (L,\,T\right ) \sim f g^{3/2} T^{-3/2} L^{-3/2}~.
\end{equation}

\subsection{First-order phase transitions}

Detailed calculations of magnetic field generation during the
electroweak and QCD phase transtions have been carried out by numerous
authors.  By and large, these groups assume that the transitions are
first-order, that is characterized by a mixed-phase regime in which
bubbles of the new phase nucleate and expand, eventually filling the
volume.  The energy associated with the bubble walls is released as a
form of latent heat.  \cite{QLS} demonstrated that a Biermann battery
can operate during the QCD phase transition.  The up, down, and
strange quarks (the three lightest quarks) have charges 2/3, -1/3, and
-1/3 respectively.  If these quarks were equal in mass, the
quark-gluon plasma would be electrically neutral.  However, the
strange quark is heavier and therefore less abundant.  The implication
is that there is a net positive charge in the quark-gluon plasma and a
net negative charge in the lepton sector.  Electric currents are
therefore generated at the bubble walls that separate the quark phase
from the baryon phase sweep through.  \cite{QLS} found that $5\,{\rm
  G}$ fields could be generated on scales of $100\,{\rm cm}$ at the
time of the QCD phase transition.  Following the arguments outlined
above, the field strength on galactic scales at the time of
recombination would be (a disappointly small) $\sim 10^{-31}\,{\rm
  G}$.

Somewhat larger estimates were obtained by \cite{CO} and \cite{SOJ}
who realized that as the hadronic regions grow, baryons would
concentrate on the bubble walls due to a ``snowplow'' effect.
(\cite{SOJ} also showed that fluid instabilities could give rise to
strong magnetic fields during the QCD phase transition.)
For reasonable parameters, they obtained fields about seven orders
of magnitude larger than those found by \cite{QLS}.

Magnetic fields can arise during cosmological phase transitions even
if they are second order --- that is, phase transitions signaled by
the smooth and continuous transition of an order parameter.  \cite{V},
for example, showed that gradients in the Higgs field vacuum
expectation value (the order parameter for the electroweak phase
transition) induce magnetic fields on a scale $\sim T_{EW}^{-1}$ with
strength of order $q_{EW}^{-1}T_{EW}^{-2}$ where $T_{EW}$ is the
temperature of the electroweak phase transition and $q_{EW}$ is the
Higgs field coupling constant.  To estimate the field on larger
scales, \cite{V} assumed that the Higgs field expectation value
executed a random walk with step size equal to the original coherence
length.  The field strengths were small ($10^{-23}\,{\rm G}$ on
$100\,{\rm kpc}$ scales) but not neglibible.

\subsection{Inverse cascade}

The discussion above suggests that strong magnetic fields are likely
to have been generated in the early Universe but that their coherence
length is so small, the effective large-scale fields are
inconsequential for astrophysics.  However, dynamical mechanisms may
lead to an increase in the coherence length of magnetic fields
produced at early times.  Chief among these is an inverse cascade of
magnetic energy from small to large scales which occurs when there is
substantial magnetic helicity \citep{FPLM}.  The effect was
investigated in the context of primordial magnetic fields (see, for
example, \cite{C, Son, FC, Brandenburg96, BJ04}.  As the Universe
expands, magnetic energy shifts to large scales as the field attempts
to achieve equilibrium while conserving magnetic helicity.  Under
suitable conditions, \cite{FC} showed that astrophysically interesting
fields with strength $~10^{-10}\,{\rm G}$ could be generated on
$10\,{\rm kpc}$ scales.

\subsection{Plasma processes capable of generating magnetic fields}
\subsubsection{Chromodynamic magnetic fields?} \label{sec:1}

The QCD regime lasts from the end of reheating until hadron formation,
roughly $t_{\mathit{ns}}\sim 10^{-6}$ s after the Big Bang (see Figure
\ref{rt-weibel-1}).  In this regime, matter comprises massive bosons,
gluons, quarks and leptons and forms a hot dense chromoplasma or
quark-gluon plasma (QGP).  At the higher temperatures, that is, not
too long after reheating, the QGP is asymptotically free and can be
considered collisionless.  Since many of the particles in the QGP
carry electric charge, under certain conditions, they can generate
induced Yang-Mills currents $j_a^\mu(x)\!=\!D_\mu F^{\mu\nu}(x)$, with
Yang-Mills field $F_{\mu \nu}= A_{\nu,\mu} - A_{\mu ,\nu} -ig [A_\mu ,
A_\nu]$ expressed through the non-Abelian gauge field
$A_{a;\nu}(x)$. The colour index $a$ corresponds to the $N^2$$-1$
colour channels. These currents couple to the electromagnetic gauge
field and consequently produce magnetic fields.  Several mechanisms
associated with the QCD phase transition were discussed in the
previous section.  Here we explore whether thermal plasma
instabilities in the QGP can lead to appreciable fields.

%In order to investigate
%thermal fluctuations on different scales, one must understand the
%response function and polarisation tensor $\Pi^{\mu\nu}(k)$ of the
%QGP, with $k\equiv(\omega,{\bf k})$ the wavenumber four-vector. In
%this section we do not intend to solve this problem; we rather state
%it as an issue that is open to future research.

%Generation of low frequency magnetic fluctuations in the QGP requires
%the presence of free energy. In which form such free energy exists is
%not known. It will cause transverse electromagnetic fluctuations that,
%in the extremely low frequency spectral range, may lead to
%quasi-stationary (on the scale of the evolution of the Universe)
%magnetic fields. Before nucleosynthesis any such field of frequency
%$\omega< 10$ MHz in the microwave (infrared) range can be considered
%stationary on the time scales prior to nucleosynthesis.

The simplest plasma mechanism is the Weibel (current filamentation)\footnote{For a
  discussion of its physics in classical plasmas see
  \cite{fried:1959}.} instability discovered in classical
plasma \citep{weibel:1959}.  Its free energy is provided by a local
pressure anisotropy $A=P_\|/P_\perp-1\neq 0$ in the
non-magnetic plasma.  $\|,\perp$ refer to the two orthogonal
directions $\hat\parallel,\hat\perp$ of the pressure tensor ${\sf
  P}=P_\perp{\sf I}+(P_\|-P_\perp){\hat\parallel}{\hat\parallel}$ whose   non-diagonal
elements are small \citep{blaizot:2002}.
% where the frequencies of plasma waves
%scale as the QCD coupling constant $\sim g$, while binary collisions
%scale as $\sim g^2$ (or even $\sim g^4$).
Pressure anisotropy creates microscopic currents and hence microscopic
magnetic fields. Free energy can also be provided by partonic beams
passing the QGP.  Such beams naturally introduce a preferred direction
may cause additional pressure anisotropy by dissipating their momentum
in some (collisionless) way.

Assuming that a cold partonic beam passes the QGP, both analytical
theory and numerical simulations %(the strong
%force is more complicated than electromagnetism!).  However,
%has been achieved in the last decade
\citep{arnold:2005,arnold:2006,arnold:2007a,arnold:2007b,arnold:2007c,rebhan:2008,romatschke:2006,schenke:2008,strickland:2006,strickland:2007}
prove that
the QCD beam-driven Weibel instability
excites magnetic fields which subsequently scatter and
thermalize the beams by magnetising the partons.
%Fast beams passing across an initially nonmagnetic chromoplasma
%give rise to conditions similar to the filamentation instability.
The linear waves, that is, oscillations of the effective quark phase
space momentum distribution ${\mathrm \Phi}_{\mathit{eff}}({ p})$ as
function of the 4-momentum, $p$, are solutions of the semi-classical
QGP dispersion relation %in a QGP plasma has been proposed by
\citep{pokrovsky:1988,mrowczynski:1988,mrowczynski:1993} in Fourier
space $k=(\omega/c,\mathbf{k})$
\begin{equation}
{\rm det}[{\bf k}^2\delta^{ij}-kk^j-\omega^2\epsilon^{ij}(|{\bf k}|)]=0
\end{equation}
with permeability (velocity ${v}^i={p}^i/\sqrt{p^lp_l}$ is the
velocity) given as functional of the effective phase space density
${\mathrm \Phi}_{\mathit{eff}}({\bf p})$, which does not depend any
more on the colour index
\begin{equation}
\epsilon^{ij}(\omega,{\bf k})=\delta^{ij}+\frac{g^2}{2\omega^2}\int\frac{d^3p}{8\pi^3}\frac{v^i[\partial\, {\mathrm \Phi}_{\mathit{eff}}({\bf p})/\partial p^l]}{\omega-{\bf k\cdot v}+i0}\left[(\omega-{\bf k\cdot v})\delta^{lj}+{k^lv^j}\right]
\end{equation}
Instability is found at low frequency, $\omega\approx 0$, non-oscillatory filamentation
modes with wave vectors, $k_\perp$,
perpendicular to the beam four-velocity, $U$.  The implication is that
stationary magnetic fields are generated.

Analytical growth rates of transverse modes, where Im\,$\omega>0$,
have been obtained for simple gaussian and other mock equilibrium
particle distributions and nuclear physics parameters.
\cite{arnold:2007a} has shown that the breakdown of perturbation
theory at momenta $p\sim g^2T$ and the fact that theory becomes
non-perturbative at high $T$ implies that it can be treated as if one
had $T=0$ plus weak coupling.  Thus, the semi-classical approach
describes long-range properties\footnote{The scalar potential $A^0$
  picks up a Debye screening mass $m_D$ and decouples at distances
  $\gg m_D^{-1}\sim(gT)^{-1}$ leaving the vector potential (transverse
  chromo-electromagnetic) fields, which is equivalent to classical
  plasmas where at frequencies $\omega>\omega_p$ above the plasma
  frequency $\omega_p$ any propagating perturbation is purely
  electromagnetic.}. In numerical simulations one takes advantage of
this fact, linearises around a stationary homogeneous locally
colourless state \citep{blaizot:2002}, and considers the evolution of
the fluctuation $W^\mu(v,x)$ of the distribution function according to
the non-Abelian Vlasov equation and Yang-Mills current density:
\begin{figure}[t!]
\centerline{{\includegraphics[width=0.7\textwidth,height=0.5\textwidth, clip=]{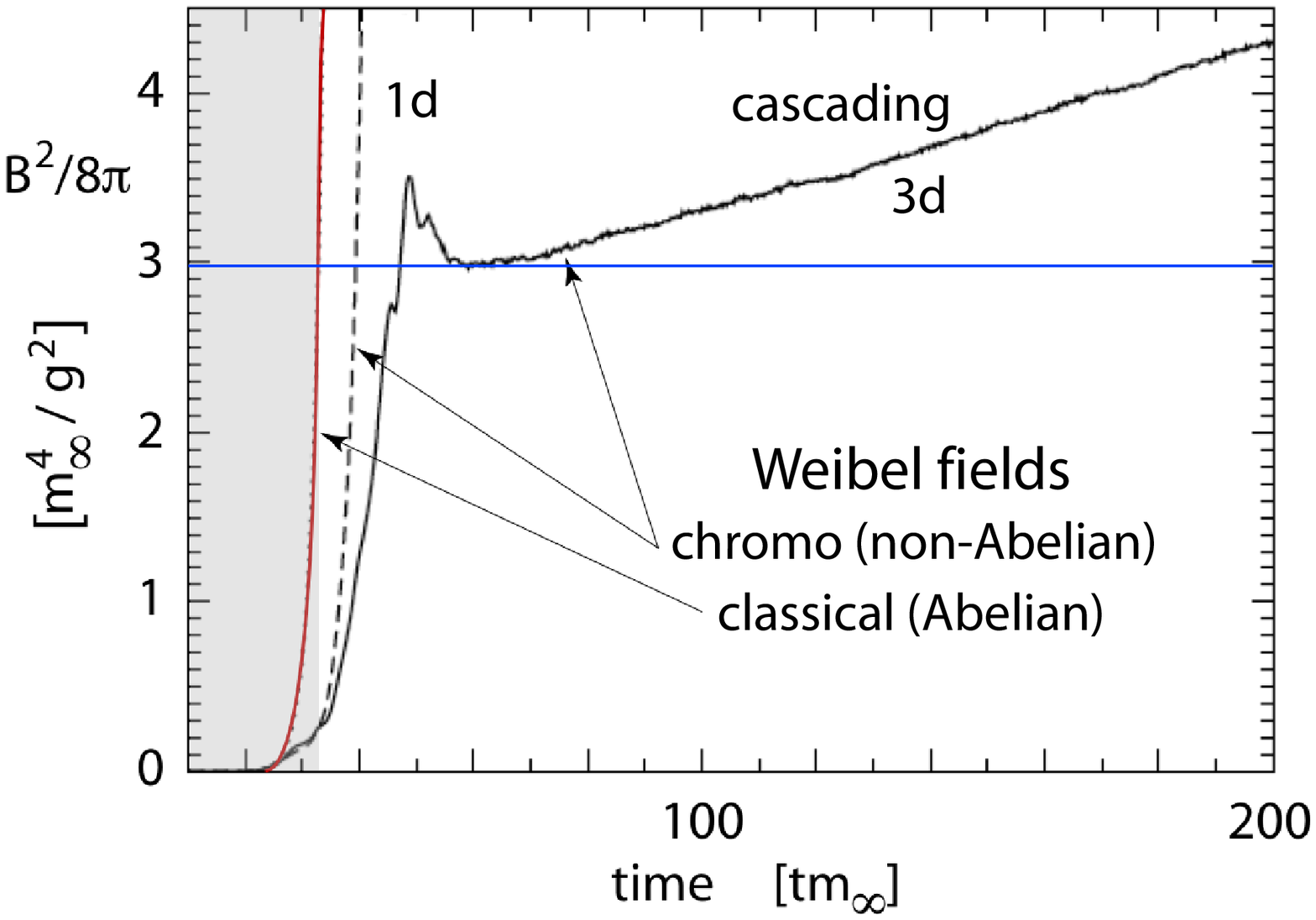} }}
\caption[ ] {\footnotesize Numerical simulation results of Weibel
fields under Abelian and non-Abelian (QGP) conditions
\citep[after][]{arnold:2005}. The Abelian and spatially 1d-cases have
a long linear phase of exponential growth. The corresponding linear
phase in the spatially 3d-case is very short (shaded region), followed
by a nonlinearly growing phase which reaches maximum field strength
and afterward increases weakly and linearly. Closer investigation has
shown that this further linear growth is caused by cascading to
shorter wavelength (larger $k$) related to increasing frequencies
thereby dispersing the free energy that feeds the nonlinear phase into
a broad spectrum of magnetic oscillations. The blue horizontal line is
the nonlinear saturation level of the low-frequency long-wavelength
Weibel magnetic field that persists during the cascade and presumably
survives it.}\label{rt-weibel-2}
\vspace{-0.3cm}
\end{figure}
\begin{equation}
v_\mu D^\mu W^\nu({\bf v},x)\!=\!-v_\sigma F^{\sigma\nu}(x), \quad j^\nu(x)\!=\!-g^2\!\!\int\frac{d^3p}{8\pi^3}\frac{p^\nu}{|{\bf p}|}\frac{\partial{\rm\Phi}_{\mathit{eff}}({\bf p})}{\partial p^\sigma}W^\sigma({\bf v},x)
\end{equation}
where $v^\nu=(1,{\mathbf p})$. The latter enters the field equations which close the system and describes the evolution of  fields and particles,
\begin{equation}
\left.
\begin{array}{ccl}
{v_\mu} D^\mu W^\nu&=&-g({\bf E+v\times B})\cdot\nabla_{\bf p}{\mathrm\Phi_{\mathit{eff}}} \\
&&\\
D_\mu F^{\mu\nu}&=&-g^2(2\pi)^{-3}{\int}{d^3p}v^\nu(D^\nu)^{-1}({\bf E+v\times B})\cdot\nabla_{\bf p}{\mathrm\Phi}_{\mathit{eff}}
\end{array}\right\}
\end{equation}
Instability requires that the effective distribution ${\mathrm\Phi}_{\mathit{eff}}({\bf p})$ is anisotropic  \citep{arnold:2005} in ${\bf p}$ to lowest order O(1). The maximum unstable wave vector $k_m$ and growth rate $\gamma_m\equiv +(\mathrm{Im}\,\omega)_m$ of the QCD Weibel mode scale as
\begin{equation}
k_m^2\sim\gamma^2_m\sim m_\infty ^2\sim g^2\int_{\mathbf p_>}{\mathrm\Phi}_{\mathit{eff}}({\mathbf p}) / |{\mathbf p}|
\end{equation}
where $m_\infty\!\sim\! g\sqrt{N_>/p_>}$ is the ``effective mass" scale defined by the spatial number density $N_>$ of the particles with momentum $p_>$ that contribute to the integral and anisotropy, i.e. $N_>\equiv \int_{\mathrm p_>}\!{\mathrm\Phi}_{\mathit{eff}}({\mathbf p})d^3p/8\pi^3$.

Figure \ref{rt-weibel-2} plots three simulation results: an Abelian
run, a spatially 1d, and a spatially
3d-non-Abelian run. Shown is the magnetic fluctuation energy density
$B^2/8\pi$ as function of time (all in proper simulation units). The
Abelian and the 1d-non-Abelian cases cover just their linear
(exponentially growing) phases. The 3d-non-Abelian case differs from
these in several important respects. Its linear phase is very short,
shown as the shaded region. It is followed by a longer nonlinear
growth phase when the magnetic energy density increases at a slower
and time dependent rate until reaching maximum, when it starts decaying.
Afterwards it recovers to end up in a further slow but now purely
algebraic linear growth.

\paragraph{Weibel saturation level. }
The nonlinear phase results from the nonlinear (wave-particle
interaction) terms that come progressively into play when the field
energy increases. The subsequent brief decay phase and the following
slow linear growth phase result from the sudden onset of a (turbulent)
long-wavelength cascade to shorter wavelength and higher frequencies.
This result has been convincingly demonstrated
\citep{arnold:2006,arnold:2007c}. The cascade transfers the excess energy that
the nonlinear instability feeds per time unit into the long wavelengths to
shorter wavelengths in such a way that in the average the energy
density in the long wavelengths (low frequencies) stays at a constant
level which, presumably, survives the cascade. This value, for the
settings of the simulation, is
\begin{equation}
{\cal E}_{\mathit{Weibel}}^{\mathit{sat}}\approx 3m_\infty^4/g^2
\end{equation}
For the coupling constant one may use \citep{bethke:2009} the
``world-average value" $g^2=4\pi\alpha_s(M_Z^2)\approx
4\pi\times0.12=0.48\pi$, where $M_Z$ is the mass of the $Z$ boson,
while for $m_\infty^2$ additional knowledge is required of the
effective parton distribution ${\mathrm\Phi}_{\mathit{eff}}$, i.e.
the state of the undisturbed distribution including the anisotropy.

Unfortunately, the above formula cannot be easily used to estimate the
long-wavelength Weibel field saturation value in the early universe.
%First, the parameters of the simulation have been adjusted to
%accelerator conditions and not to those of the early universe.  Indeed,
%the conditions in the early Universe, especially in view of any
%possible thermal anisotropy or the presence of parton beams, are
%poorly known, though it seems reasonable to assume that
%inhomogeneities do exist.  These question, both for the early Universe
%and for terrestrial accelerator experiments remain open areas of
%active research.
If we accept that the saturation level is
robust within few orders of magnitude, then, because
$m_\infty^2\sim\omega_p^2/\sqrt{\theta}$, one has as for an estimate
(in physical units)
\begin{equation}
B^\mathit{sat}\approx \omega_p^2\sqrt{2\mu_0\hbar/c^3\theta}
\end{equation}
 where $\omega_p$ is the effective chromo-plasma frequency $\omega_p^2=gT/m_D\lambda_D^2$ which is related to the temperature $T$, Debye mass $m_D$, and screening length $\lambda_D$, and $\theta\sim \tan^{-1}(A^{-1})$ is an effective anisotropy angle \citep{arnold:2007c}.
  With these numbers one estimates quite a strong saturation magnetic field
 \begin{equation}
 B^\mathit{sat}\approx10^{-19}(m_e/M_Z\sqrt{\theta})n_\mathit{eff}\sim2\times10^{-25}
 n_\mathit{eff} /\sqrt{\theta}\quad{\mathrm G}
 \end{equation}
a value that depends linearly on the effective parton density
$n_\mathit{eff}$. Taking, say,
$n_\mathit{eff}\sim10^{10}$ cm$^{-3}$, it yields a large QCD saturated
low-frequency seed magnetic field of the order of
$B^\mathit{sat}\sim10^{-5}\mu$G. This seems
unrealistically high. What it shows, however, is that only very small
anisotropies $\theta\sim$\,O(1) are required for production of
magnetic fields during the QGP phase even if the above estimate is
wrong by several orders of magnitude. It seems thus worthy of further
considering this possibility of primordial magnetic field generation.

The mass density in the standard model varies by about 40 orders of
magnitude from reheating until quark-hadron transition where it is
$n_m^\mathit{qht}\approx 10^{20}$\,kg/m$^3$. At weak unification it is
of the order of $n_m^\mathit{ew}\approx 10^{33}$\,kg/m$^3$. Number
densities depend on the dominant particle mass chosen. If, for
simplicity, we assume a mass $M\sim 10^3$\,GeV, then number densities
at the quark-hadron transition are still of the order of
$n^\mathit{qht}_\mathit{eff}\sim 10^{21}$\,cm$^{-3}$. With such high
effective densities one obtains extreme values for the saturated
magnetic field the order of $B^\mathit{sat}\sim10^{-4}$ G and even
larger at an earlier time. This is probably unrealistic and thus could
hardly be right. Clearly, if such high fields existed they must have
been attenuated during further evolution (expansion) of the universe
to the low values needed after recombination ended.

\paragraph{Thermal fluctuation level. }
The argument might be weakened when no anisotropy exists at all. Then
one is led to the determination of the thermal magnetic Weibel
fluctuation level which, of course, is well below the saturation
level estimated above. For its estimation the effective {\it isotropic
equilibrium} distribution ${\mathrm\Phi}_{\mathit{eff}}({\mathbf p})$
and the complex response function, i.e. the QGP polarisation tensor
$\Pi^{\mu\nu}$, are needed whose determination requires solution of the Vlasov equation.
A rough
estimate of the thermal level can be found from the above simulation results by
taking advantage of the evolution equation of the average magnetic
energy density $\langle B^2(t)\rangle$ in  long wavelength
fluctuations and assuming that the magnetic energy density measured at
a certain time $t_1$ in the linear phase only evolved from the thermal
level $\langle b^2(t=0)\rangle$ at time $t=0$ according to
 \begin{equation}
\langle B^2(t_1)\rangle\simeq \langle b^2(t=0)\rangle \exp\,2\gamma_mt
\end{equation}
>From the linear phase in Figure \ref{rt-weibel-2} one finds that
$\gamma_m\approx 0.28/m_\infty$. This value used in the last equation
yields an approximate (initial) thermal Weibel magnetic energy density
level of $\langle b^2\rangle\approx
2.88\times10^{-8}m_\infty^4g^{-2}$, which corresponds to a thermal
Weibel magnetic fluctuation of average amplitude
\begin{equation}
\langle b^\mathit{th}_\mathit{Weibel} \rangle \sim 3.4\times 10^{-29}n_\mathit{eff}\quad {\mathrm G}
\end{equation}
during the corresponding QCD phase; at the quark-hadron transition
this becomes $\langle
b^\mathit{th}_\mathit{qht}\rangle\!\sim\!\!100\,\mu$G, which still is
very large for thermal background fields. Levels such high as the
estimate would, probably, be subject to attenuation in the classical
plasma phase before decoupling if they should account for the current
large scale fields. Of course, this value can only give a hint on the
possible importance of thermal fluctuations as it has been taken and
rescaled from the available simulations which have been performed for
other purposes (thermalisation of nuclear matter beams). Whether
realistic or not can be decided only after developing a QCD theory for
the cosmological QGP phase.\begin{figure}[t!]
\centerline{{\includegraphics[width=0.6\textwidth,height=0.5\textwidth,
clip=]{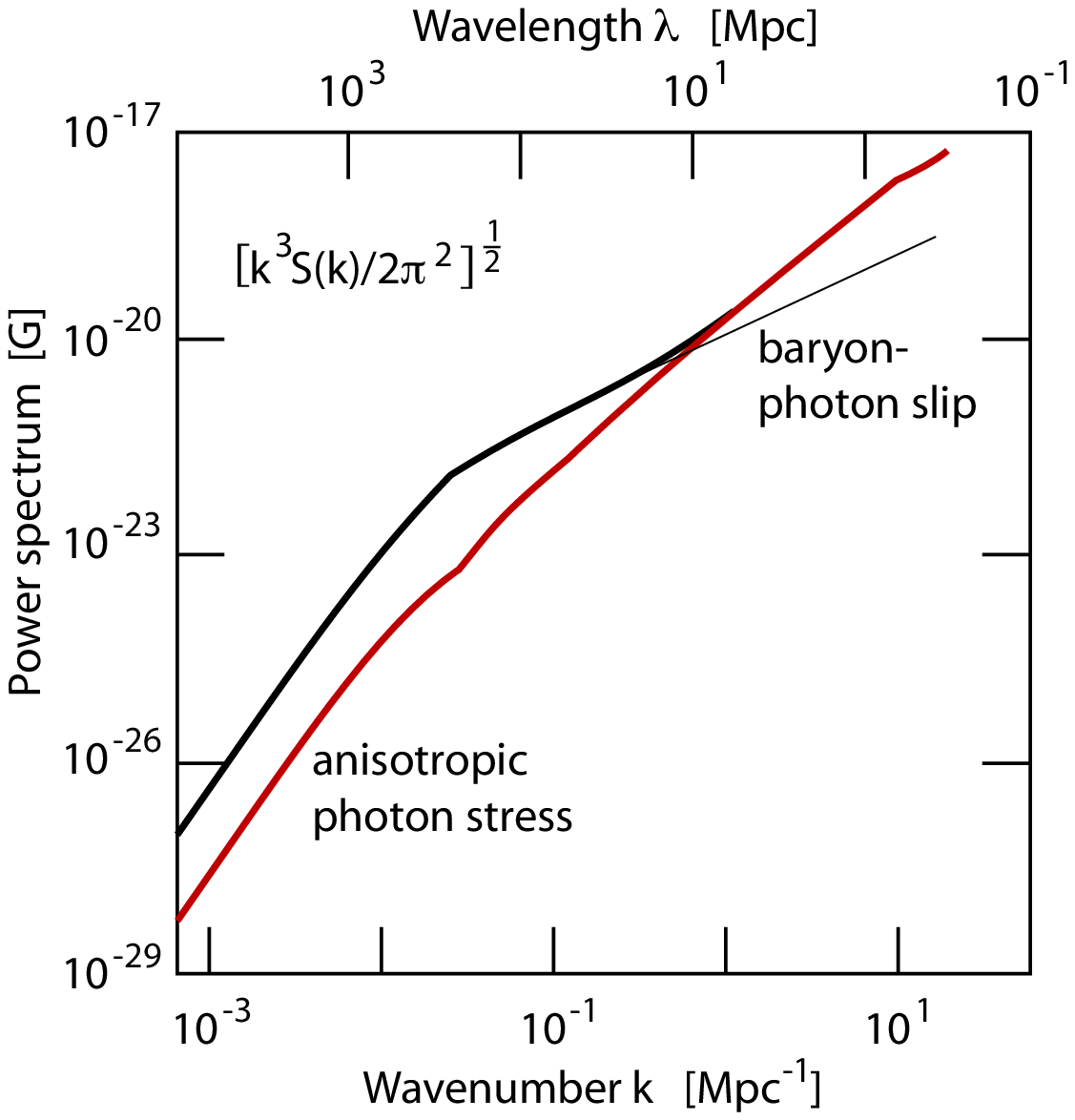} }}
\caption[ ] {\footnotesize The spectrum $S(k)$ of magnetic fields in
the pre-recombination era generated from cosmological density
perturbations \citep[after][]{ichiki:2006} plotted in units of
magnetic field (G) and shown to be composed of contributions from
baryon-phon slip and anisotropic photon stresses. Below roughly
$k<$1/Mpc the baryon slip dominates the spectrum while for larger $k$
(shorter wavelength than roughly 10 Mpc) the anisotropic stresses
contribute most. }\label{rt-weibel-3}
\vspace{-0.3cm}
\end{figure}

\subsubsection{Thermal fluctuations shortly before photon-matter decoupling.}

The previous section dealt with the possible generation of low
frequency seed magnetic fields during the QCD phase of the early
universe.  A more conventional proposal has been elaborated recently
\citep{ichiki:2006,ichiki:2007} and is proposed to work during the
classical plasma phase before recombination. These authors make the
reasonable assumption that during this epoch the coupling between
photons and electrons by Compton scattering is much stronger than the
coupling between photons and ions. Cautious examination of the
particle-photon interaction in a cosmological density fluctuation
field indeed shows that pressure anisotropy and currents are
induced. The generalised Ohm's law that includes the photon
interaction allows for a finite electron current flow because of the
differences in the bulk electron and proton velocities caused. This
effect produces the desired magnetic fields. However, only the second
order density perturbations in the Compton scattering terms lead to
fields that survive on a range of spatial scales. The power spectrum
$S(k)$ of the fields in the long wavelength range (see Figure
\ref{rt-weibel-3}) scales as $\sqrt{k^3S(k)}\propto k$ Here the
photon-caused anisotropic stress dominates.

Seed magnetic fields reach values of $B sim10^{-18}$ G on scales of
$\sim$1 Mpc and $B \sim10^{-14}$ G on $\sim$10 kpc scales.  After
decoupling these fields decay adiabatically with expansion of the
universe. In a standard cosmology they should today be of strength
$B(t_0)\sim 10^{-24}$ G at 1 Mpc and $\sim10^{-20}$ G at 10 kpc and
may have played a role in structure formation after recombination.

%This conventional mechanism is well suited to
%explain the existence of magnetic fields on the largest scales in the
%universe as resulting from a primordial source that is acting
%relatively late right before recombination and decoupling of matter
%from photons. The largest of the scales become amplified when they
%re-enter the horizon thus being signals of the amplification processes
%that acted at the horizon at the time of re-entry \cite{ichiki:2007}.

%%%%%%%%%%%%%%%%%%%%%%%%%%%%%%%%%%%%%%%%%%%%%%%%%%%%%%%%%%%%%%%%%%%%%%%

\section{Magnetic fields and the cosmic microwave background}

The field of cosmology has witnessed a revolution lead, in large part,
by detailed measurements of the CMB anisotropy and polarization
spectra.  Fundamental cosmological parameters such density parameters
of baryons, dark matter, and dark energy and the Hubble constant are
now known to a precision unimaginable just two decades ago.

If magnetic fields were present at the time of matter-radiation
decoupling or soon after, then they would have an effect on the
anisotropy and polarization of the CMB
(see \cite{subramanian06,durrer07} for reviews).
First, a very large scale (effectively homogeneous) field
would select out a special direction, lead to
anisotropic expansion around this direction,
hence leading to a quadrupole anisotropy
in the cosmic microwave background (CMB)
\citep[see, for example,][]{Thorne}.
The degree of isotropy of the CMB then implies a limit
of several nG on the strength of such a field redshifted to the
current epoch \citep{BFS97}.

Primordial magnetogenesis scenarios on the other hand
generally lead to tangled fields, plausibly Gaussian random,
characterized by say a spectrum $M(k)$.
This spectrum is normalized by giving the field strength
$B_0$, at some fiducial scale, and as measured at the present epoch,
assuming it decreases with expansion as $B=B_0/a^2(t)$, where $a(t)$
is the expansion factor.
Since magnetic and radiation energy densities both scale with
expansion as $1/a^4$, we can characterize the magnetic field effect
by the ratio $B_0^2/(8\pi\rho_{\gamma 0}) \sim 10^{-7} B_{-9}^2$
where $\rho_{\gamma 0}$ is the present day energy density in radiation, and
$B_{-9} = B_0/(10^{-9} G)$.
Magnetic stresses are therefore small compared to
the radiation pressure for nano Gauss fields.

Nevertheless, the scalar, vector and tensor parts of the
perturbed stress tensor associated with primordial magnetic fields
lead to corresponding metric perturbations, including gravitational
waves. Further the compressible part of the Lorentz force leads to
compressible (scalar) fluid velocity and associated density
perturbations, while its vortical part leads to
vortical (vector) fluid velocity perturbation.
These magnetically induced metric and velocity perturbations lead
to both large and small angular scale anisotropies in the CMB temperature
and polarization.

The scalar contribution has been the most subtle
to calculate, and has only begun to be understood by several groups
\citep{Giovanni_Kunze,Yamazaki08,Finelli08,SL10}.
The anisotropic stress
associated with the magnetic field leads in particular to the possibility
of two types of scalar modes, a potentially dominant mode
which arises before neutrino decoupling
sourced by the magnetic anisotropic stress.
And a compensated mode which remains after the growing neutrino
ansiotroic stress has compensated the magnetic anisotropic
stress (cf. \cite{SL10} for detailed discussion).
The magnetically induced compressible fluid perturbations,
also changes to the acoustic peak structure of the
angular anisotropy power spectrum \citep[see, for example,][]{Adams}.
However, for nano Gauss fields, the CMB anisotropies due to the magnetized
scalar mode are grossly subdominant to the anisotropies generated by
scalar perturbations of the inflaton.

Potentially more important is the contribution of the
Alfv\'en mode driven by the rotational component of
the Lorentz force \citep{SB98,mack02,SSB03,lewis04}.
Unlike the compressional mode, which gets
strongly damped below the Silk scale, $L_{S}$
due to radiative viscosity
\citep{Silk68}, the Alfv\'en mode behaves like an over damped oscillator.
This is basically because the phase velocity of oscillations,
in this case the Alfv\'en velocity, is
$V_{A} \sim 3.8 \times 10^{-4} c B_{-9}$ much smaller
than the relativistic sound speed $c/\sqrt{3}$.
Note that for an over damped oscillator there is one normal mode which
is strongly damped and another where the velocity starts from
zero and freezes at the terminal velocity till the damping becomes
weak at a latter epoch. The net result is that
the Alfv\'en mode survives Silk damping down to much smaller scales;
$L_A \sim (V_A/c) L_{S} \ll L_{S}$, the
canonical Silk damping scale \citep{Jedamzik98,SB98a}.
The resulting baryon velocity leads to a CMB temperature
anisotropy, $\Delta T \sim 5 \mu K (B_{-9}/3)^2$
for a scale invariant spectrum, peaked below
the Silk damping scale (angular wavenumbers $l > 10^3$).

The magnetic anisotropic stress also induces
tensor perturbations, resulting in a comparable CMB
temperature anisotropy, but now peaked on large
angular scales of a degree or more \citep{DFK00}.
Both the vector and tensor perturbations lead
to ten times smaller B-type polarization anisotropy,
at respectively small and large angular scales
\citep{Seshadri01,SSB03,mack02,lewis04}. Note that
inflationary generated scalar perturbations only produce
the E-type mode. The small
angular scale vector contribution in particular can potentially
help to isolate the magnetically induced
signals \citep{SSB03}.

A crucial difference between the magnetically
induced CMB anisotropy signals compared to those
induced by inflationary scalar and tensor perturbations,
concerns the statistics associated with the signals.
Primordial magnetic fields lead to non-Gaussian statistics
of the CMB anisotropies even at the lowest order,
as magnetic stresses and the temperature anisotropy they induce
depend quadratically on the magnetic field.
In contrast, CMB non-Gaussianity due to inflationary scalar
perturbations arises only as a higher order effect.
A computation of the nongaussianity of the magnetically induced
signal has begun \citep{SS09,caprini09,cai10},
based on earlier calculations of non-Gaussianity in the
magnetic stress energy \citep{BC05}.
This new direction of research
promises to lead to tighter constraints or a detection of
strong enough primordial magnetic fields.

A primordial magnetic field leads to a number of other effects
on the CMB which can probe its existence:
Such a field in the inter galactic medium
can cause Faraday rotation of the polarized component
of the CMB, leading to the generation of new B-type
signals from the inflationary E-mode signal \citep{KL}.
Any large-scale helical component of the field leads to
a parity violation effect, inducing non-zero T-B and E-B
cross-correlations \citep{KR05}; such cross-correlations between
signals of even and odd parity are neccessarily zero
in standard inflationary models. The damping
of primordial fields in the pre-reombination era can
lead to spectral distortions
of the CMB (Jedamzik, Katalinic \& Olinto 2000),
while their damping in the post-recombination
era can change the ionization and thermal history of the
universe and hence the electron
scattering optical depth as a function of redshift
(see \cite{Sethi05, Tashiro06a, Schleicher08}, and
  Section 5.3).
Future CMB probes like PLANCK
can potentially detect the modified CMB anisotropy signal from
such partial re-ionization.
In summary primordial magnetic fields of a few nG lead
to a rich variety of effects on the CMB and thus are
potentially detectable via observation of CMB anisotropies.

%As reviewed in \cite{GR},
%there are four distinct effects: anisotropy induced by very large
%scale (effectively homogeneous) field \citep[see, for example,][]{Thorne};
%changes to the acoustic peak structure of the
%%angular anisotropy power spectrum \citep[see, for example,][]{Adams};
%damping of baryon-photon sound waves by MHD effects \citep{JKO};
%and polarization effects due to Faraday rotation \citep{KL}.  We also
%note that the CMB further provides constraints due to the
%reionization optical depth which changes in the presence of strong
%magnetic fields (see \cite{Sethi05, Tashiro06a, Schleicher08}, and
%  Section 5.3).

\section{Implications of strong primordial fields in the post-recombination universe}

If strong magnetic fields have been produced during phase-transitions
in the early universe, and if these fields had some non-zero helicity,
they may have remained strong until recombination and beyond
\citep{Christensson01, BJ04}. They could then affect the thermal
and chemical evolution during the dark ages of the Universe, the
formation of the first stars, and the epoch of reionization.

\subsection{Implications during the dark ages}\label{darkages}

At high redshift $z>40$, the universe is close to homogeneous, and the
evolution of the temperature, $T$, is governed by the competition of
adiabatic cooling, Compton scattering with the CMB and, in the
presence of strong magnetic fields, ambipolar diffusion. It is thus
given as
\begin{eqnarray}
\frac{dT}{dz}&=&\frac{8\sigma_T a_R T_{\sm{rad}}^4}{3H(z)(1+z)m_e c}\frac{x_e\,(T-T_{\sm{rad}})}{1+f_{\fHeI}+x_e}\nonumber\\
&+&\frac{2T}{1+z}-\frac{2(L_{\sm{AD}}-L_{\sm{cool}})}{3nk_B H(z)(1+z)},\label{temp}
\end{eqnarray}
where $L_{\sm{AD}}$ is the heating function due to ambipolar diffusion
(AD), $L_{\sm{cool}}$ the cooling function \citep{Anninos97},
$\sigma_T$ the Thomson scattering cross section, $a_R$ the
Stefan-Boltzmann radiation constant, $m_e$ the electron mass, $c$ the
speed of light, $k_B$ Boltzmann's constant, $n$ the total number
density, $x_e=n_\fe/n_H$ the electron fraction per hydrogen atom,
$T_{\sm{rad}}$ the CMB temperature, $H(z)$ is the Hubble factor and
$f_{\fHeI}$ is the number ratio of \HeI and \HI nuclei.

AD occurs due to the friction between ionized and neutral species, as only the former are directly coupled to the magnetic field. Primordial gas consists of several neutral and ionized species, and for an appropriate description of this process, we thus adopt the multi-fluid approach of \citet{Pinto08a}, defining the AD heating rate as
\begin{equation}
 L_{\sm{AD}}=\frac{\eta_{\sm{AD}}}{4\pi}\left|\left(\nabla\times\vec{B}\right)\times\vec{B}/B\right|^2,\label{ambiheat}
\end{equation}
where $\eta_{\sm{AD}}$ is given as
\begin{equation}
 \eta_{\sm{AD}}^{-1}=\sum_n \eta_{\sm{AD},n}^{-1}.\label{etaAD}
\end{equation}
In this expression, the sum includes all neutral species $n$, and
$\eta_{\sm{AD},n}$ denotes the AD resistivity of the neutral species
$n$. We note that the AD resistivities themselves are a function of magnetic field strength, temperature and chemical composition.

In the primordial IGM, the dominant contributions to the total
resistivity are the resistivities of atomic hydrogen and helium due to
collisions with protons.  These are calculated based on the momentum
transfer coefficients of \citet{Pinto08b}. As the power spectrum of
the magnetic field is unknown, we estimate the expression in
Eq.~(\ref{ambiheat}) based on the coherence length $L_B$ , given as
the characteristic scale for Alfv{\'e}n damping \citep{Jedamzik98,
  Subramanian98, Seshadri01}. Contributions from decaying MHD
turbulence may also be considered, but are negligible compared to the
AD heating \citep{Sethi05}.

The additional heat input provided by AD affects the evolution of the
ionized fraction of hydrogen, $x_\fpp$, which is given as
\begin{eqnarray}
\frac{dx_\fpp}{dz}&=&\frac{[x_e x_p n_H \alpha_H -\beta_H (1-x_p)e^{-h_p\nu_{H, 2s}/kT}] }{H(z)(1+z)[1+K_H(\Lambda_H+\beta_H)n_H(1-x_p)]}\nonumber\\
&\times&[1+K_H\Lambda_H n_H(1-x_p)]-\frac{k_{\sm{ion}}n_H x_p}{H(z)(1+z)}.\label{ion}
\end{eqnarray}
Here, $n_H$ is the number density of hydrogen atoms and ions, $h_p$
Planck's constant, $k_{\sm{ion}}$ is the collisional ionization rate
coefficient \citep{Abel97}. Further details of notation, as well as
the parametrized case B recombination coefficient for atomic hydrogen
$\alpha_H$, are given by \citet{Seager99}. The chemical evolution of
the primordial gas is solved with a system of rate equations for the
chemical species \HMd, \HzIId, \HzId, \HeHIId, \DId, \DIId, \DMd,
\HDIId and \HDI based on the primordial rate coefficients tabulated by
\citet{Schleicher08}. For the mutual neutralization rate of \HM and
\HIId, we use the more recent result of \citet{Stenrup09}. The
evolution of the magnetic energy density $E_B=B^2/8\pi$ is given as
\begin{equation}
\frac{dE_B}{dt}=\frac{4}{3}\frac{\partial\rho}{\partial t}\frac{E_B}{\rho}-L_{\sm{AD}}.\label{bfield}
\end{equation}
The first term describes the evolution of the magnetic field {in a
homogeneous universe in the absence of specific magnetic energy
generation or dissipation mechanisms.  } The second term accounts for
corrections due to energy dissipation via AD.

The dynamical implications of magnetic fields can be assessed from the magnetic Jeans mass, the critical mass scale for gravitational forces to overcome magnetic pressure. In the large-scale IGM, it is given as
 \citep{Subramanian98, Sethi05}
\begin{equation}
M_J^B\sim10^{10}M_\odot\left(\frac{B_0}{3\ \mathrm{nG}} \right)^3.\label{mJeans}
\end{equation}
Due to ambipolar diffusion, strong magnetic fields also affect the gas temperature and thus the thermal Jeans mass, i.e. the critical mass scale required to overcome gas pressure. It is defined as
\begin{equation}
%M_J=2M_\odot \left(\frac{c_s}{0.2\ \mathrm{km/s}} \right)^3\left(\frac{n}{10^3\ \mathrm{cm}^{-3}} \right)^{-1/2}.\label{thJeans}
M_J=\left( \frac{4\pi\rho}{3} \right)^{-1/2}\left(\frac{5k_B T}{2\mu G m_P}\right)^{3/2}\label{thJeans}
\end{equation}
with Boltzmann's constant $k_B$, the mean molecular weight $\mu$ and the proton mass $m_p$.

To describe virialization in the first minihalos, we employ the
spherical collapse model of \citet{Peebles93} for pressureless dark matter until an
overdensity of $\sim200$ is reached. Equating cosmic time with the timescale from the spherical collapse model allows one to calculate the overdensity $\rho/\rho_b$ as a function of time or redshift. In this model, we further assume that the formation of the protocloud will reduce the coherence length of the magnetic field to the size of the cloud.

\begin{figure}[t]
\includegraphics[scale=0.5]{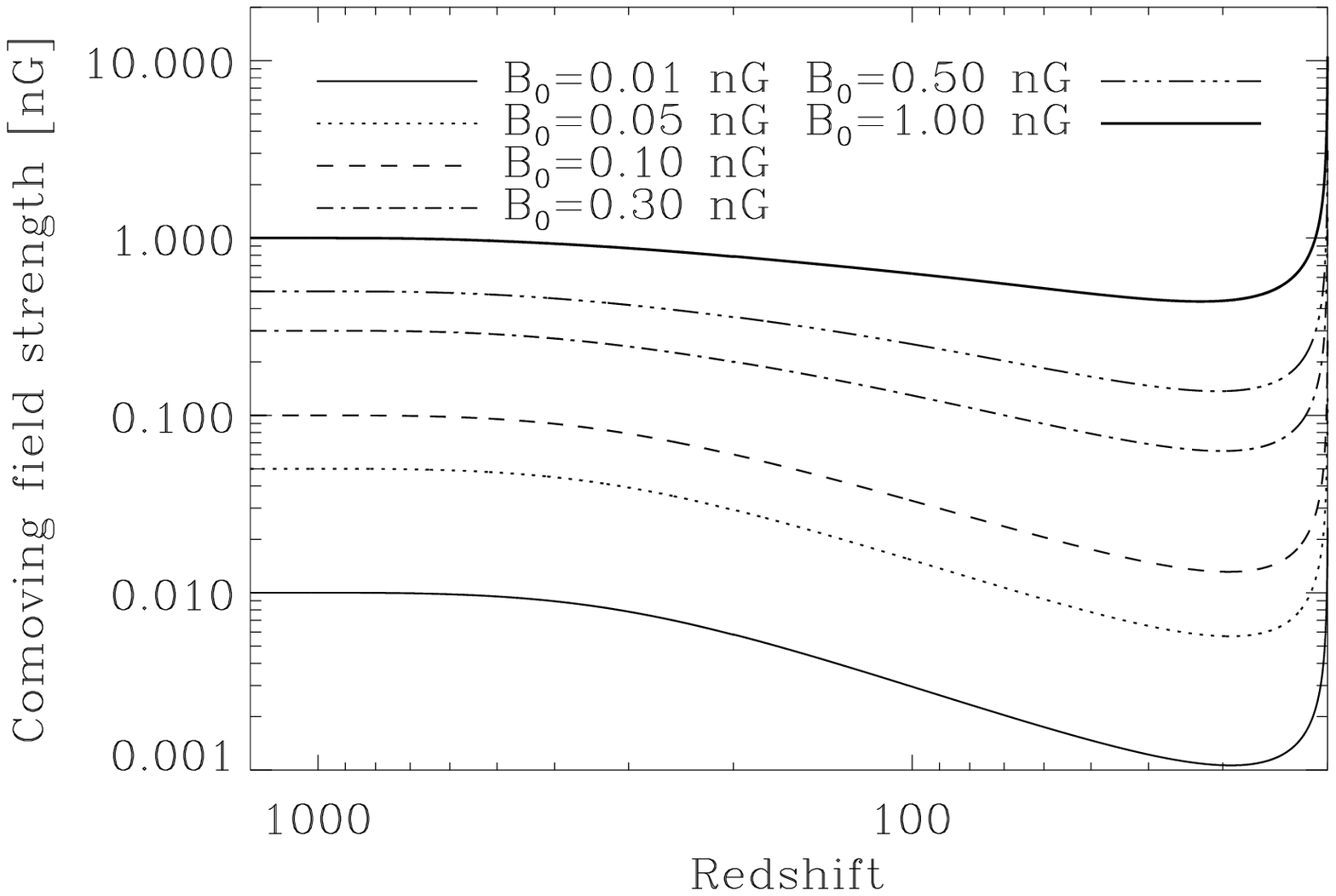}
\caption{The evolution of the comoving magnetic field strength due to AD as a function of redshift for different initial comoving field strengths, from the homogeneous medium at $z=1300$ to virialization at $z=20$.}
\label{fig:bfieldIGM}
\eef

 \begin{figure}[t]
 \includegraphics[scale=0.5]{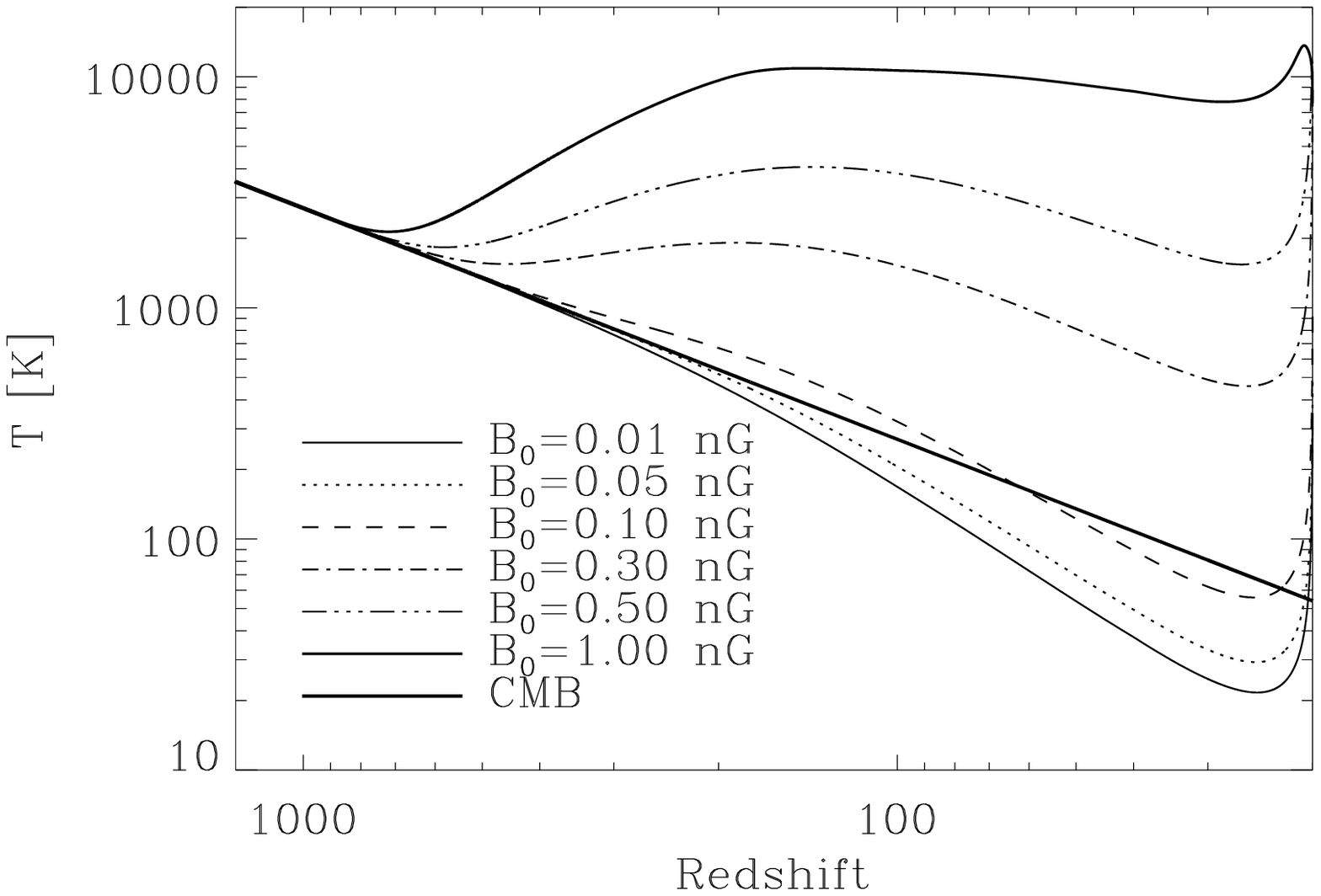}
 \caption{The gas temperature evolution in the IGM as a function of redshift for different comoving field strengths, from the homogeneous medium at $z=1300$ to virialization at $z=20$. For the case with $B_0=0.01$~nG, we find no difference {in the thermal evolution} compared to the zero-field case. }
 \label{fig:tempIGM}
 \eef

\begin{figure}[t]
\includegraphics[scale=0.5]{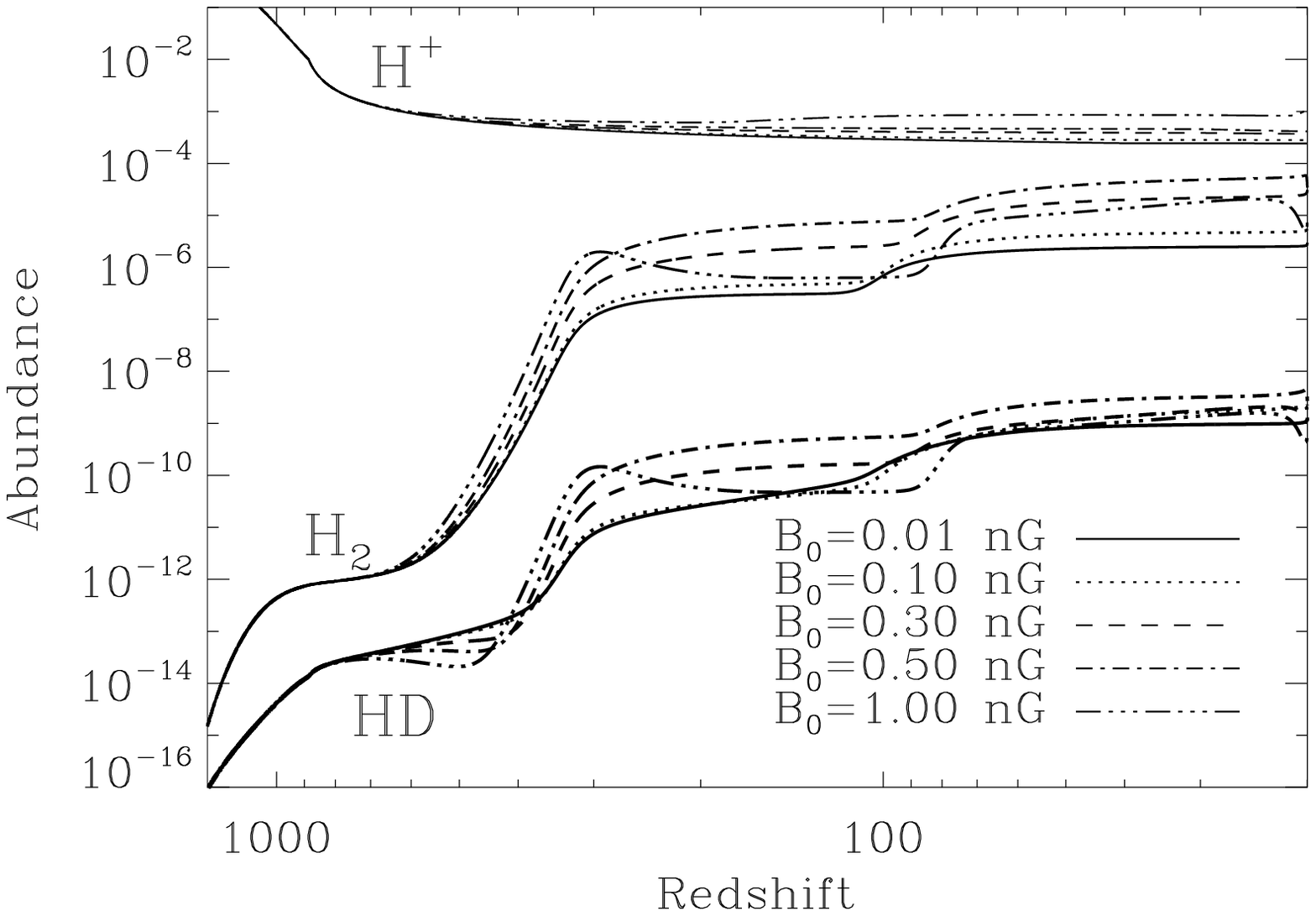}
\caption{The evolution of ionization degree, \HzI and \HDI abundances as a function of redshift for different comoving field strengths, from the homogeneous medium at $z=1300$ to virialization at $z=20$. For the case with $B_0=0.01$~nG, we find no difference {in the chemical evolution} compared to the zero-field case.}
\label{fig:ionIGM}
\eef

The evolution of the magnetic field strength, the IGM temperature and
the chemical abundances of different species have been calculated by
\citet{Schleicher09prim} using an extension of the recombination code
RECFAST \citep{Seager99}. The results are shown in
Figs.~\ref{fig:bfieldIGM}-\ref{fig:ionIGM}. As shown in
Fig.~(\ref{fig:bfieldIGM}), ambipolar diffusion primarily affects
magnetic fields with initial comoving field strengths of $0.2$~nG or
less. For stronger fields, the dissipation of only a small fraction of
their energy increases the temperature and the ionization fraction of
the IGM to such an extent that AD becomes less effective. For comoving
field strengths up to $\sim0.1$~nG, the additional heat from ambipolar
diffusion is rather modest and the gas in the IGM cools below the CMB
temperature due to adiabatic expansion. However, it can increase
significantly for stronger fields and reaches $\sim10^4$~K for a
comoving field strength of $1$~nG, where Lyman $\alpha$ cooling and
collisional ionization become efficient and prevent a further increase
in temperature. The increased temperature enhances the ionization
fraction and leads to larger molecule abundances at the onset of star
formation.

\subsection{Implications for the formation of the first stars}\label{collapse}

We now explore in more detail the consequences of magnetic fields for
the formation of the first stars, during the protostellar collapse
phase. For this purpose, a model describing the chemical and thermal
evolution during free-fall collapse was developed by
\citet{Glover09} and extended by \citet{Schleicher09prim} for the
effects of magnetic fields. Particularly important with
respect to this application is the fact that it correctly models the evolution of
the ionization degree and the transition at densities of
$\sim10^8$~cm$^{-3}$ where Li$^+$ becomes the main charge
carrier. Based on the Larson-Penston type self-similar solution
\citep{Larson69, Penston69, Yahil83}, we evaluate how the collapse
timescale is affected by the thermodynamics of the gas.

During protostellar collapse, magnetic fields are typically found to
scale as a power-law with density $\rho$. Assuming ideal MHD with flux
freezing and spherical symmetry, one expects a scaling with
$\rho^{2/3}$ in the case of weak fields. Deviations from spherical
symmetry such as expected for dynamically important fields give rise
to shallower scalings, e.g. $B \propto \rho^{0.6}$ \citep{Banerjee06},
$B \propto \rho^{1/2}$ \citep{Hennebelle08a, Hennebelle08}. Based on numerical simulations of
\citet{Machida06}, we find an empirical scale law
\begin{equation}
 \alpha=0.57\left(\frac{M_J}{M_J^B} \right)^{0.0116}.\label{fit}
\end{equation}
In a collapsing cloud, the more general expression for the magnetic Jeans mass,
\begin{equation}
 M_J^B=\frac{\Phi}{2\pi\sqrt{G}},\label{mgJeans}
 \end{equation}
 is adopted. In this prescription, $\Phi=\pi r^2 B$ denotes the
 magnetic flux, $G$ the gravitational constant, $r$ an appropriate
 length scale. The calculation of the magnetic Jeans mass thus
 requires an assumption regarding the size of the dense
 region. Numerical hydrodynamics simulations show that they are
 usually comparable to the thermal Jeans length \citep{Abel02,
   Bromm04}. This is also suggested by analytical models for
 gravitational collapse \citep{Larson69, Penston69, Yahil83}.To
 account for magnetic energy dissipation via AD, we calculate the AD
 heating rate from Eq.~(\ref{ambiheat}) and correct the magnetic field
 strength accordingly. We note that due to the large range of
 densities during protostellar collapse, additional processes need to
 be taken into account to calculate the AD resistivity correctly. In
 particular, at a density of $\sim10^9$~cm$^{-3}$, the three-body \HzI
 formation rates start to increase the \HzI abundance significantly,
 such that the gas is fully molecular at densities of
 $\sim10^{11}$~cm$^{-3}$. As a further complication, the proton
 abundance drops considerably at densities of $\sim10^8$~cm$^{-3}$,
 such that Li$^+$ becomes the main charge carrier \citep{Maki04,
   Glover09}. These effects are incorporated in our multi-fluid
 approach.

\begin{figure}[t]
\includegraphics[scale=0.5]{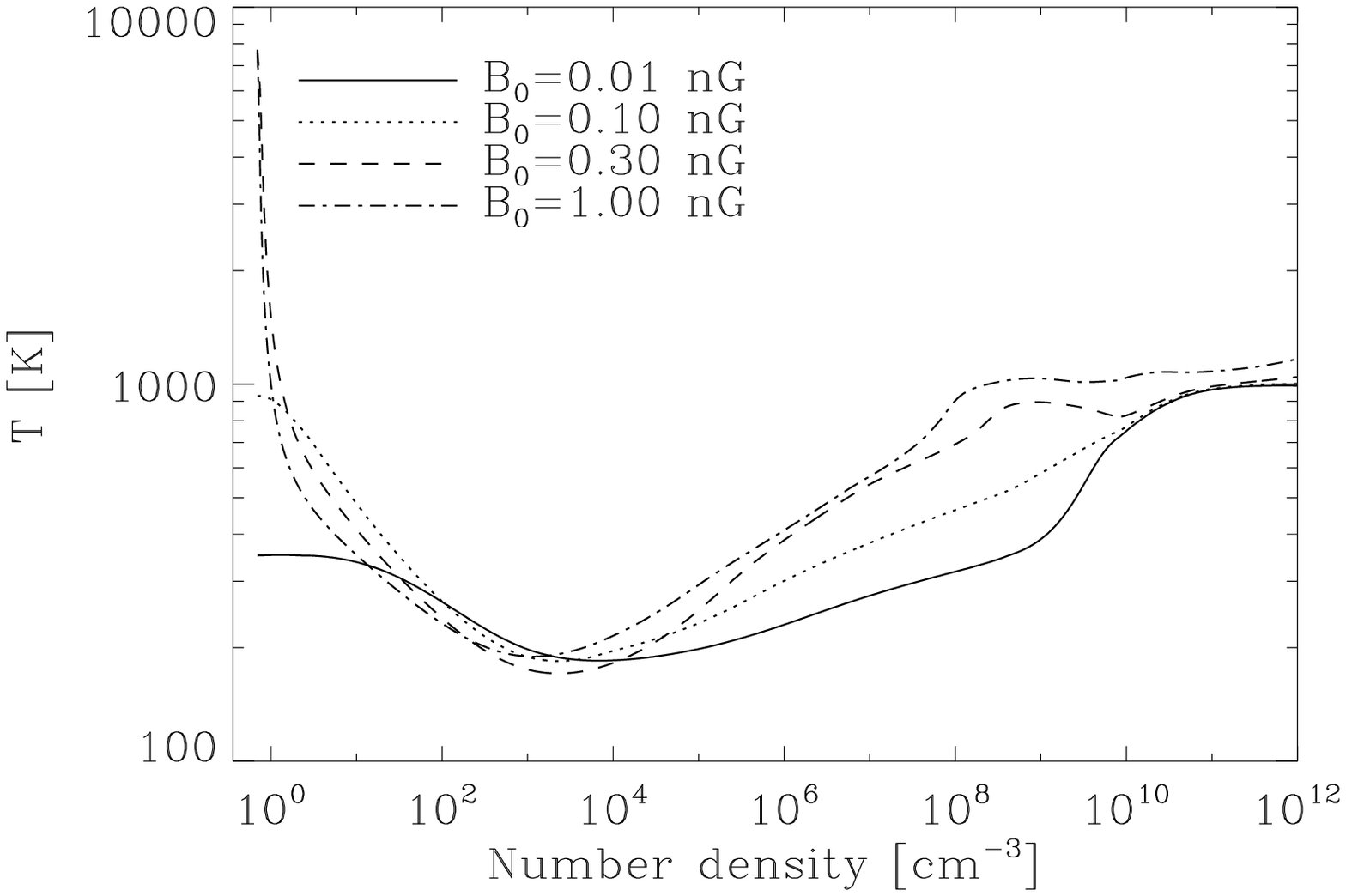}
\caption{The gas temperature as a function of density for different comoving field strengths. For $B_0=0.01$~nG, the thermal evolution corresponds to the zero-field case.}
\label{fig:tempcoll}
\eef

%\begin{figure}[t]
%\includegraphics[scale=0.5]{f7.eps}
%\caption{Ionization degree, \HzI and \HDI abundance as a function of density for different comoving field strengths. For $B_0=0.01$~nG, the thermal evolution corresponds to the zero-field case. }
%\label{fig:abuncoll}
%\eef

\begin{figure}[t]
\includegraphics[scale=0.5]{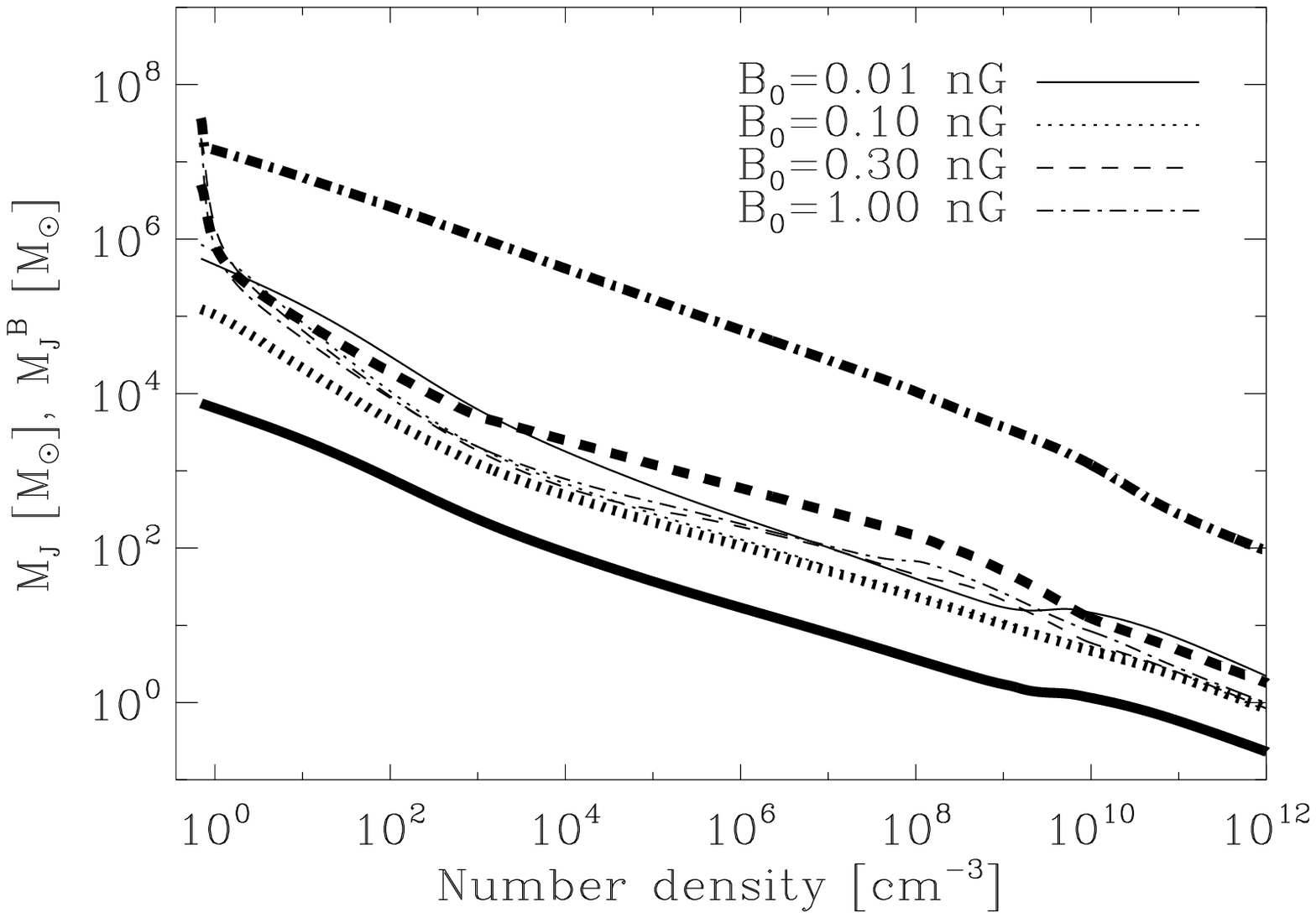}
\caption{Thermal (thin lines) and magnetic (thick lines) Jeans mass as a function of density for different comoving field strengths.  For $B_0=0.01$~nG, the thermal evolution and thus the thermal Jeans mass corresponds to the zero-field case.}
\label{fig:jeanscoll}
\eef

As an initial condition for these model calculations, we use the physical field
strength and the chemical abundances obtained from the spherical
collapse model in section~\ref{darkages}. The relation between co-moving and physical field
strength at the beginning of this calculation is thus given in
Table~\ref{tab:models}.

\begin{table}[htdp]
\begin{center}
\begin{tabular}{cc}
$B_0$~[nG] & $B$~[nG]  \\
\hline
$1$  &  $4.8\times10^3$ \\
$0.3$ & $6.3\times10^2$ \\
$0.1$ & $1.3\times10^2$ \\
$0.01$ & $1.0\times10^1$
\\
\hline
\end{tabular}
\end{center}
\caption{The physical field strength $B$ at beginning of collapse as a function of the comoving field strength $B_0$ used to initialize the IGM calculation at $z=1300$. %The labels in the figure plots refer to the initial comoving field strength.
}
\label{tab:models}
\end{table}%

Fig.~\ref{fig:tempcoll} shows the calculated temperature evolution as a function
of density for different comoving field strengths. For comoving fields
of $0.01$~nG or less, there is virtually no difference in the
temperature evolution from the zero-field case. For comoving fields of
$\sim0.1$~nG, cooling wins over the additional heat input in the early
phase of collapse, and the temperature decreases slightly below the
zero-field value at densities of $10^3$~cm$^{-3}$. At higher
densities, the additional heat input dominates over cooling and the
temperature steadily increases. At densities of $\sim10^9$~cm$^{-3}$,
the abundance of protons drops considerably and increases the AD
resistivity defined in Eq.~(\ref{etaAD}) and the heating rate until
Li$^+$ becomes the main charge carrier. In particular for comoving
fields larger than $\sim0.1$~nG, this transition is reflected by a
small bump in the temperature evolution due to the increased heating
rate in this density
range.

%The evolution of the abundances of free electrons and the molecules \HzI and \HDI are
%given in Fig.~\ref{fig:abuncoll}. With higher initial field strength,
%the abundance of free electrons increases, inducing the formation of
%more molecules. Interestingly, the molecular fraction becomes
%independent of field strength at densities of $\sim10^{10}$~cm$^{-3}$
%where three-body \HzI formation takes
%over.

Apart from the transition where Li$^+$ becomes the dominant charge
carrier, the magnetic field strength usually increases more rapidly
than $\rho^{0.5}$, and weak fields increase more rapidly than strong
fields. This is what one naively expects from Eq.~(\ref{fit}), and it
is not significantly affected by magnetic energy dissipation. Another
important point is that comoving fields of only $10^{-5}$~nG are
amplified to values of $\sim1$~nG at a density of
$10^3$~cm$^{-3}$. Such fields are required to drive protostellar
outflows that can magnetize the IGM \citep{Machida06}.

Fig.~\ref{fig:jeanscoll} shows the evolution of the thermal and
magnetic Jeans mass during collapse. The thermal Jeans masses are
quite different initially, but as the temperatures reach the same
order of magnitude during collapse, the same holds for the thermal
Jeans mass. The thermal Jeans mass in this late phase has only a weak
dependence on the field strength. As expected, the magnetic Jeans
masses are much more sensitive to the magnetic field strength, and
initially differ by about two orders of magnitude for one order of
magnitude difference in the field strength. For comoving fields of
$\sim1$~nG, the magnetic Jeans mass dominates over the thermal one and
thus determines the mass scale of the protocloud. For $\sim0.3$~nG,
both masses are roughly comparable, while for weaker fields the
thermal Jeans mass dominates. The magnetic Jeans mass shows features
both due to magnetic energy dissipation, but also due to a change in
the thermal Jeans mass, which sets the typical length scale and thus
the magnetic flux in the case that $M_J>M_J^B$.

The uncertainties in these models have been explored further by
\citet{Schleicher09prim}, and an independent calculation including
stronger magnetic fields has been provided by \citet{Sethi10}. We also note that the consequences of initially weak magnetic fields for primordial star formation are explored in more detail in the next chapter.

\subsection{Implications for reionization}
Strong magnetic fields may influence the epoch of reionization in
various ways. As discussed above, they may affect the formation of the
first stars and change their mass scale, and thus their feedback
effects concerning cosmic reionization and metal enrichment. They may
further give rise to fluctuations in the large-scale density field and
potentially enhance high-redshift structure formation \citep{Kim96,
  Sethi05, Tashiro06a}. On the other hand, the increased thermal and
magnetic pressure may indeed suppress star formation in small halos
and thus delay reionization \citep{Schleicher08, Rodrigues10}. In both
cases, unique signatures of the magnetic field may become inprinted in
the 21~cm signature of reionization, which may help to constrain or
detect such magnetic fields with LOFAR\footnote{LOFAR homepage:
  http://www.lofar.org/}, EDGES\footnote{EDGES homepage:
  http://www.haystack.mit.edu/ast/arrays/Edges/} or SKA\footnote{SKA
  homepage: http://www.skatelescope.org/} \citep{TashiroSugiyama06a,
  Schleicher09a}. Upcoming observations of these facilities may thus
provide a unique opportunity to probe high-redshift magnetic fields in
more detail.

\section{Seed fields in the post-recombination Universe}

We have seen that magnetic fields arise naturally during inflation and
during phase transitions after inflation but before recombination.
However, the effective field strength on galactic scales may well be
exceedingly small.  Indeed, the seed fields for the galactic dynamo
may well arise from astrophysical processes rather than exotic
early-Universe ones.  In this section, we review three processes which
can generate magnetic field in the post-recombination Universe.

\subsection{Biermann battery}

In the hierarchical clustering scenario, proto-galaxies acquire
angular momentum from tidal torques produced by neighboring systems
\citep{Hoyle, Peebles69, W}.  However, these purely gravitational
forces cannot generate vorticity (the gravitational force can be
written as the gradient of a potential whose curl is identially zero)
and therefore the existence of vorticity must arise from gasdynamical
processes such as those that occur in shocks.  More specifically,
vorticity is generated whenever one has crossed pressure and density
gradients.  In an ionized plasma, this situation drives currents
which, in turn, generate magnetic field.  This mechanism, known as the
Biermann battery and originally studied in the context of stars
\citep{B} was considered in the cosmological context by \cite{PS,
KCOR, DW} and \citet{xu08}.  A simple order of magnitude estimate yields
\begin{equation}
B_{\rm biermann} ~\simeq~ \frac{m_p c}{e}\omega
~\simeq ~ 3\times 10^{-21}\,\left (\frac{\omega}{\rm km s^{-1}\,kpc^{-1}}
\right ) \,{\rm Gauss}
\end{equation}
where $\omega$ is the vorticity.  Since the vorticity in the solar
neighborhood is of order $30\,{\rm km\, s^{-1}\,kpc^{-1}}$ we expect
seed fields of order $10^{-19}\,{\rm G}$.

\subsection{First-generation stars}

The first generation of stars provides another potential source of
seed fields for the galactic dynamo.  Even if stars are born without
magnetic fields, a Biermann battery will generate weak fields which
can then be rapidly amplified by a stellar dynamo.  If the star
subsequently explodes or loses a significant amount of mass through
stellar winds, the fields will find their way into the interstellar
medium and spread throughout the (proto) galaxy.  Simple estimates by
\cite{Syr} illustrate the viability of the mechanism.  There have been
some $10^8$ supernovae over the lifetime of the galaxy, each of which
spreads material through a $\left (10\,\rm pc\right )^3$ volume.
Using values for the field strength typical of the Crab nebula, one
therefore expects the galaxy to be filled by $10\,\rm pc$ regions with
field strengths $\sim 3\,\mu G$.  Assuming the same $L^{-3/2}$ scaling
discussed above, one finds a field strength of $10^{-11}\,{\rm G}$ on
$10\,{\rm kpc}$ scales, a value significantly larger than the ones
obtained by more exotic early Universe mechanisms.

The strong amplification of seed magnetic fields during primordial
star formation has been suggested in a number of works.  Analytical
estimates by \cite{PS, Tan04} and \cite{Silk06} suggest that large-scale
dynamos as well as the magneto-rotational instability could
significantly amplify weak magnetic seed fields until saturation.
Even before, during the protostellar collapse phase, the small-scale
dynamo leads to an exponential growth of the magnetic fields, as found
in both semi-analytic and numerical studies \citep{Schleicher10, Sur10}.

\subsection{Active galactic nuclei}

Strong magnetic fields almost certainly arise in active galactic
nucleii (AGN).  These fields will find their way into the
intergalactic medium via jets thereby providing a potential source of
magnetic field for normal galaxies \citep{Hoyle69, Rees87, Rees94,
  DL}.  The potential field strengths due to this mechanism may be
estimated as follows: The rotational energy associated with the
central compact (mass $M$) object which powers the AGN can be
parametrized as $fMc^2$ where $f < 1$.  If we assume equipartition
between rotational and magnetic energy within a central volume $V_c$,
we find
\begin{equation}
B_c = \left (\frac{8\pi fMc^2}{V_c}\right )^{1/2}
\end{equation}
If this field then expands adiabatically to fill a ``galactic'' volume
$V_g$ one finds $B_g = B_c\left (V_c/V_g\right )^{2/3}$.
\cite{Hoyle69} considered values $M=10^9\,M_\odot$, $f=0.1$,
$V_g\simeq \left (100\,{\rm kpc}\right )^3$ and found $B_c\simeq
10^9\,{\rm G}$ and $B_g\simeq 10^{-5}\,{\rm G}$.

\section{Conclusions}

The origin of the seed magnetic fields necessary to prime the galactic
dynamo remains a mystery and one that has become more, rather than
less, perplexing over the years as observations have pushed back the
epoch of microgauss galactic fields to a time when the Universe was a
third its present age.  Numerous authors have explored the possible
that the galactic fields observed today and at intermediate redshift
have their origin in the very early Universe.

The impetus for the study of early Universe magnetic fields came from
the successful marriage of ideas from particle physics and cosmology
that occurred during the latter half of the last century.  It is a
remarkable prediction of modern cosmology that the particles and
fields of the present-day Universe emerged during phase transitions a
fraction of a second after the Big Bang.  The electromagnetic and weak
interactions became distinct during the electroweak phase transitions
at $t\simeq 10^{-12}\,{\rm s}$ while baryons replaced the quark-gluon
plasma at $t\simeq 10^{-6}\,{\rm s}$.  Perhaps more fantastical is the
notion that galaxies, clusters, and superclusters arose from
quantum-produced density perturbations that originated during
inflation at even earlier times.  This idea is supported by strong
circumstantial from the CMB anisotropy spectrum, so much so, that it
is now part of the standard lore of modern cosmology.

Both inflation and early Universe phase transitions have many of the
ingredients necessary for the creation of magnetic fields.  If our
understanding of inflation-produced density perturbations is correct,
the similar quantum fluctuations in the electromagnetic field will
lead to fields on the scales of galaxies and beyond.  As well,
electromagnetic currents, and hence fields, will almost certainly be
driven during both the electroweak and quark-hadron phase transitions.

Early Universe schemes for magnetic field generation, however, face
serious challenges.  In the standard electromagnetic theory, and in a
flat or closed FLRW cosmology, inflation-produced fields at diluted by
the expansion to utterly negligible levels.  One may address this
issue by considering fields in an open Universe and ``just-so''
inflation scenario, that is, a scenario with just enough e-folds of
inflation to solve the flatness problem.  Alternatively, one may
incorporate additional couplings of the field to gravity into the
theory though many terms lead to unwanted consequences which render
the theory unphysical.  Observations and advances in theoretical
physics may settle the issue.  For example, if future determinations
find that the Universe has a slight negative curvature (density
parameter for matter and dark energy slightly less than one), it would
give some credance to the idea of superadiabatic field amplification
in an open Universe.  On the other hand, string theory may point us to
couplings between gravity and electromagnetism that naturally generate
fields during inflation.

The main difficulty with fields generated from phase transitions
arises from the small Hubble scale in the very early Universe.  Strong
fields are almost certainly produced by one of a number of mechanism.
But the coherence length is so small, the effective field strength on
galactic scales is likely to be well-below the level of interest for
astrophysics.  An inverse cascade may help; if the field has a net
helicity, the magnetic field energy will be efficiently transferred
from small to large scales.  But even if fields are uninteresting for
the galactic dynamo, they may have an effect on processes in the
post-recombination Universe such as reionization and the formation of
the first generation of stars.

Where, if not the early Universe, did the seed fields for the galactic
dynamo arise?  Astrophysics provides a number of promising
alternatives.  Galactic disks have angular momentum {\it and}
vorticity.  While the former is generated by the gravitational
interaction between neighboring protogalaxies, the latter comes form
gasdynamical effects which necessarily generate magnetic fields via
the Biermann battery.  As well, magnetic fields, rapidly created and
amplified inside some early generation of stars or in active galactic
nucleii, can be dispersed throughout the intergalactic medium.

%\bibliographystyle{aps-nameyear}      % basic style, author-year citations
%\bibliography{example}   % name your BibTeX data base
%\nocite{*}

\begin{acknowledgements}

  The work by DR is supported in part by by National Research
  Foundation of Korea through grant KRF-2007-341-C00020.  LMW is
  supported by a Discovery Grant from the Natural Sciences and
  Engineering Research Council of Canada.  DS receives funding from
  the European Community's Seventh Framework Programme
  (/FP7/2007-2013/) under grant agreement No 229517.  He further
  thanks the Landesstiftung Baden-W{\"u}rttemberg via their program
  International Collaboration II for financial support.

\end{acknowledgements}

\end{document}